\documentclass[prb,preprint,showpacs,preprintnumbers,amsmath,amssymb]{revtex4}

\usepackage{graphicx}
\usepackage{dcolumn} 
\usepackage{bm}

\newcommand{\gl}{{\langle\!\langle}}
\newcommand{\gr}{{\rangle\!\rangle}}

\newcommand{\imag}{{\rm \,Im\,}}

\begin{document}
\title{Decoupling method for dynamical mean field theory calculations}

\author{Harald O. Jeschke}
\email{jeschke@physics.rutgers.edu}
\affiliation{Department of Physics \& Astronomy, Rutgers University, 
136 Frelinghuysen Road, NJ 08854-8019, USA}
\author{Gabriel Kotliar}
\email{kotliar@physics.rutgers.edu}
\affiliation{Department of Physics \& Astronomy, Rutgers University, 
136 Frelinghuysen Road, NJ 08854-8019, USA}

\date{\today}

\begin{abstract}
In this paper we explore the use of an equation of motion decoupling
method as an impurity solver to be used in conjunction with the
dynamical mean field self-consistency condition for the solution of
lattice models. We benchmark the impurity solver against exact
diagonalization, and apply the method to study the infinite $U$
Hubbard model, the periodic Anderson model and the $pd$ model. This
simple and numerically efficient approach yields the spectra expected
for strongly correlated materials, with a quasiparticle peak and a
Hubbard band. It works in a large range of parameters, and therefore can
be used for the exploration of real materials using LDA+DMFT.
\end{abstract}

% 71.27.+a Strongly correlated electron systems; heavy fermions
% 71.30.+h Metal-insulator transitions and other electronic transitions

\maketitle

\section{Introduction}
Dynamical mean field theory (DMFT) was developed over the past 15
years into a powerful tool for the treatment of strongly correlated
electron systems~\cite{kotliar:96,metzner:89,kotliar:92}. DMFT is
based on the idea of mapping a complicated lattice model onto a single
impurity model coupled to a noninteracting bath. It relies on the
observation that the self energy $\Sigma({\bf k},i\omega_n)$ becomes ${\bf
k}$-independent in infinite dimensions
$d=\infty$~\cite{muellerhartmann:89}, making a single site treatment with
only temporal fluctuation exact in this limit. The DMFT approach
derives its strength from the fact that it becomes exact in this
nontrivial limit of $d=\infty$ or infinite lattice coordination. Perhaps
surprisingly, DMFT proves to be a very good approximation even in
$d=3$ dimensions. By replacing complicated models with a single
impurity model, the DMFT equations can then be solved with one of the
methods that have been developed to solve the Anderson impurity model.

The study of correlated materials has until a few years ago been
conducted with two approaches that are very different in spirit. On
the one hand, density functional theory (DFT) calculations in the
local density approximation (LDA) has proven invaluable in the
determination of the electronic structure of real materials but there
are a number of strongly correlated materials where its predictions
are even in qualitative disagreement with experiment. On the other
hand, the study of model Hamiltonians has provided qualitative
understanding of may systems with strong correlations but due to its
dependence on parameters this method lacks predictive power for new
materials. The combination of the two approaches in the form of
LDA+DMFT~\cite{kotliar:97} promises to deepen our understanding of
strongly correlated materials as some initial successes
demonstrate~\cite{kotliar:01,held:01,held:03}.

DMFT reduces the quantum many body problem to a single site problem,
namely an impurity model in a medium, and requires the solution of an
Anderson impurity model for arbitrary values of the bath. When the
LDA+DMFT is carried out self-consistently in a multiband situation,
the impurity model has to be solved many times, and becomes the
bottleneck of the LDA+DMFT algorithm. Therefore it is important to
find impurity solvers that are reliable and computationally cheap.
Currently, the usual choices for solving the Anderson impurity model
in the framework of LDA+DMFT are quantum Monte Carlo (QMC), the
non-crossing approximation (NCA), and the iterated perturbation theory
(IPT). Nevertheless, each of these methods has some drawbacks limiting
its range of applicability. The QMC method is essentially exact, but
becomes prohibitively expensive at low temperatures and for high
interaction strength $U$.  The NCA approximation, applied to the
impurity model, exceeds the unitarity limit at low temperatures and
leads to pathologies in the solutions of the DMFT equations. The
extension of the IPT scheme, a method which was very successful at
arbitrary filling in the one orbital situation, has encountered
difficulties in its extension to the multiorbital case. This provides
the motivation of this article to investigate the usefulness of a
previously known decoupling scheme in the context of DMFT.

The method for the solution of the Anderson impurity model proposed
here aims at working with an arbitrary noninteracting DOS as
input. Nevertheless, we intend to show that even for the solution of
model Hamiltonians like, {\it e.$\,$g.} the Hubbard Hamiltonian, a
DMFT scheme with a closed set of equations gained from a decoupling
scheme is superior to the direct solution of that Hamiltonian with
decoupling methods.

\section{Theory}
%\subsection{The impurity solver}
The method of writing equations of motion (EOM) for the Anderson
impurity model and decoupling them in order to close the system of
equations has a long history~\cite{theumann:69,lacroix:81,meir:93}. In
the derivation of the integral equation for the solution of the
infinite $U$ Anderson impurity model we follow the approach and the
decoupling scheme of T. Costi~\cite{costi:86}. The Hamiltonian for a
mixed valent impurity is~\cite{costi:86}
\begin{equation}
H=\sum_{kn}\varepsilon_kc^+_{kn}c_{kn} + \sum_n E_{fn}X^{nn}+E_{f0}X^{00} + \sum_{kn}\bigl( V_{kn}^*c^+_{kn}X^{0n}+V_{kn}X^{n0}c_{kn}\bigr)
\label{eq:infiniteUhamiltonian}\end{equation}
We determine the equation of motion for the $m$-channel f-electron
Green's function (GF) $F_m(\omega)=\gl X^{0m};X^{m0} \gr$ by writing
\begin{equation}
 \omega \gl X^{0m};X^{m0} \gr = \langle[X^{0m},X^{m0}]_+\rangle + \gl [X^{0m},H];X^{m0} \gr 
\end{equation}
and evaluating the appearing (anti-) commutators. The result is
\begin{equation}\begin{split}
(\omega -\varepsilon_f)\gl X^{0m};X^{m0} \gr=& \langle X^{00}+X^{mm}\rangle + \sum_kV_k\gl (X^{00}+X^{mm})c_{km} ;X^{m0}\gr \\&+ \sum_{\substack{k\\n\neq m}}V_k\gl X^{nm}c_{kn} ;X^{m0} \gr\,,
\label{eq:gf1}\end{split}\end{equation}
assuming that the hybridization $V_k$ does not depend on the $z$
component $m$ of the angular momentum $J$. The abbreviation $\varepsilon_f\equiv
E_{fm}-E_{f0}$ was introduced. The averages over the $X$ operators are
$\langle X^{00}\rangle=1-n_f$ and $\langle X^{mm}\rangle=n_f/N$ where the total number of $f$
electrons is calculated as
\begin{equation}
n_f=-\frac{N}{\pi}\int d\omega'f(\omega')F''_m(\omega')\,,
\label{eq:nf}\end{equation}
with the notation $F_m(\omega)=F'_m(\omega)+i F''_m(\omega)$.  For the higher order
Green's functions on the rhs of Eq.~\eqref{eq:gf1} we also write
equations of motion:
\begin{align}
(\omega - \varepsilon_k)\gl (X^{00}+X^{mm})c_{kn} ;X^{m0}\gr &= \label{eq:hgf1}\\ V_k \gl X^{0m};X^{m0} \gr 
+  \sum_{\substack{q\\n\neq m}} V_q\gl &c^+_{qn}X^{0n}c_{km};X^{m0} \gr 
+  \sum_{\substack{q\\n\neq m}} V_q\gl c_{qn}X^{n0}c_{km};X^{m0} \gr\notag\\
(\omega-\varepsilon_k)\gl X^{nm}c_{kn} ;X^{m0} \gr&=\label{eq:hgf2}\\ 
-\langle X^{n0}c_{kn}\rangle + \sum_qV_q&\gl X^{n0}c_{qm}c_{kn};X^{m0} \gr 
+\sum_qV_q\gl X^{0m}c^+_{qn}c_{kn};X^{m0} \gr \notag
\end{align}
We now employ the decoupling scheme already given in
Ref.~\cite{costi:86} that conserves particle number and angular
momentum ($n\neq m$ is assumed):
\begin{align}
\gl c^+_{qn}X^{0n}c_{km};X^{m0} \gr &\simeq \langle c^+_{qn}X^{0n} \rangle \gl c_{km};X^{m0} \gr\\
\gl c_{qn}X^{n0}c_{km};X^{m0} \gr &\simeq \langle c_{qn}X^{n0} \rangle \gl c_{km};X^{m0} \gr\\
\gl X^{n0}c_{qm}c_{kn};X^{m0} \gr &\simeq \langle c_{kn}X^{n0} \rangle \gl c_{qm};X^{m0} \gr\label{eq:decoup3}\\
\gl X^{0m}c^+_{qn}c_{kn};X^{m0} \gr &\simeq \langle c^+_{qn}c_{kn} \rangle \gl X^{0m};X^{m0} \gr
\label{eq:decoupling}\end{align}
Note a sign difference in Eq.~\eqref{eq:decoup3} with respect to
Ref.~\cite{costi:86}.  The Greens function $\gl c_{qm};X^{m0} \gr$
appearing here can again be determined from its equation of motion
\begin{equation}\begin{split}
\gl c_{qm};X^{m0} \gr &= \frac{V_q}{\omega - \varepsilon_q} \gl X^{0m};X^{m0} \gr
\label{eq:hgf3}\end{split}\end{equation}
This leads to the equation from which the $f$ electron Greens function can be determined:
\begin{equation}
F_m(\omega) = \frac{1-n_f+\frac{n_f}{N} +I_1(\omega)}{\omega -\varepsilon_f- \Delta(\omega) +I_2(\omega) -\Delta(\omega)I_1(\omega)}\,,
\label{eq:gf3bI}\end{equation}
with the sums over correlation functions
\begin{align}
I_1(\omega) &= -\sum_{\substack{k\\n\neq m}}\frac{V_k}{\omega - \varepsilon_k}\langle X^{n0}c_{kn}\rangle\label{eq:int1}\,,\\
I_2(\omega) &= -\sum_{\substack{kq\\n\neq m}} \frac{V_kV_q}{\omega - \varepsilon_k}\langle c^+_{qn}c_{kn} \rangle\label{eq:int2}\,,
\end{align}
and the hybridization function
\begin{equation}
\Delta(\omega)= \sum_k\frac{V_k^2}{\omega - \varepsilon_k}\,.
\label{eq:delta}\end{equation}
In the degenerate models we study in this article, the sums over $n$
with $n\neq m$ simply lead to factors of $N-1$. The sums over $k$ and $q$
can be simplified further.

To that end, we replace the correlation functions by integrals over
the imaginary part of the corresponding Greens function
\begin{equation}\begin{split}
\langle c^+_{kn} c_{qn} \rangle = -\frac{1}{\pi}\int d\omega' f(\omega') \imag\gl c_{qn};c^+_{kn} \gr\,.
\label{eq:correlation}\end{split}\end{equation}
The conduction electron Greens function $\gl c_{qn};c^+_{kn} \gr$ is
determined from its equation of motion
\begin{equation}\begin{split}
\gl c_{qn};c^+_{kn} \gr &= \frac{\delta_{kq}}{\omega - \varepsilon_q} + \frac{V_kV_q}{(\omega - \varepsilon_k)(\omega - \varepsilon_q)}\gl X^{0n};X^{n0}\gr
\label{eq:exactccgf}\end{split}\end{equation}
Now in order to simplify the sums in Eqs.~\eqref{eq:int1} and \eqref{eq:int2}, we employ the identity
\begin{equation}
\frac{1}{(\omega - \varepsilon)(\omega' - \varepsilon)}=\frac{1}{\omega' - \omega}\biggl[\frac{1}{\omega - \varepsilon}-\frac{1}{\omega' - \varepsilon}\biggr]\,.
\label{eq:identity}\end{equation}
This allows us to identify occurrences of the hybridization function \eqref{eq:delta}, and we find
\begin{align}
I_1(\omega) &=\frac{N\!-\!1}{\pi}\int d\omega'\frac{f(\omega')}{\omega' - \omega}\Bigl[ F''_m(\omega')\Delta(\omega)-\imag\bigl\{ F_m(\omega') \Delta(\omega') \bigr\}\Bigr]\label{eq:int1final}\\
I_2(\omega) &=\frac{N\!-\!1}{\pi} \int d\omega'\frac{f(\omega')}{\omega' - \omega}\Bigl[-\Delta''(\omega')+ \Delta(\omega)\imag\bigl\{ F_m(\omega')\Delta(\omega')\bigr\} - \imag\bigl\{ F_m(\omega')\Delta(\omega')^2\bigr\}  \Bigr]\label{eq:int2final}
\end{align}
Eqs.~\eqref{eq:gf3bI} together with \eqref{eq:int1final},
\eqref{eq:int2final} and the definition of $n_f$ \eqref{eq:nf} form an
integral equation for $F_m(\omega)$ that can be solved iteratively. In
order to compute the integrals of Eqs.~\eqref{eq:int1final} and
\eqref{eq:int2final} we introduce the following real functions:
\begin{equation}\begin{split}
A_m(\omega) &= -f(\omega)\imag F_m(\omega)\\
B(\omega) &= f(\omega)\imag \Delta(\omega)\\
C_m(\omega) &= f(\omega)\imag \bigl\{ F_m(\omega) \Delta(\omega)\bigr\} \\
D_m(\omega) &= f(\omega)\imag \bigl\{ F_m(\omega) \Delta(\omega)^2\bigr\} 
\end{split}\end{equation}
Now the integrals read
\begin{equation}\begin{split}
I_1(\omega) &=\frac{1}{\pi}\sum_{\substack{n\neq m}} \biggl[\Delta(\omega) \int \frac{d\omega'}{\pi} \frac{A_m(\omega')}{\omega - \omega'} +\int \frac{d\omega'}{\pi} \frac{C_m(\omega')}{\omega - \omega'}\biggr]\\
I_2(\omega) &=\frac{1}{\pi}\sum_{\substack{n\neq m}} \biggl[\int \frac{d\omega'}{\pi}\frac{B(\omega')}{\omega - \omega'}-\Delta(\omega) \int \frac{d\omega'}{\pi} \frac{C_m(\omega')}{\omega - \omega'} +\int \frac{d\omega'}{\pi} \frac{D_m(\omega')}{\omega - \omega'} \biggr]
\end{split}\end{equation}
Thus, the calculation of the integrals reduces to simple evaluation of
Kramers Kronig integrals. The imaginary part for example of the first
such integral is $-i\pi A_m(\omega')$.

I turns out that this set of equations on the real frequency axis is
solved easily for the Anderson impurity model, but as we add self
consistency conditions in order to solve more complicated models in
the DMFT approximation, convergence depends strongly on a good initial
guess of the solution. For this purpose, we write equations analogous
to Eqs.~\eqref{eq:gf3bI}, \eqref{eq:int1final}, \eqref{eq:int2final}
and \eqref{eq:nf} on the Matsubara axis. Matsubara Greens functions
are much more smooth than their counterparts on the real frequency
axis and thus converge more easily. Nevertheless, the calculation of
the Greens function on the imaginary axis does not make the real axis
calculation redundant: Firstly, the analytic continuation to the real
axis is only accurate for low frequencies due to a lack of high
frequency information in the Matsubara Greens function. Secondly, the
dependence of the imaginary frequency grid on temperature
\begin{equation}
i \omega_n=(2 n+1)\pi T
\label{eq:iwn}\end{equation}
means that at high temperatures, the low frequency part of the Greens
function is very badly resolved, while at very low temperatures, an
inordinate number of imaginary frequencies is necessary to describe
the Greens function for all frequencies for which it significantly
differs from zero. This means that from a practical point of view, the
Matsubara Greens function is best calculated at an intermediate
temperature, providing via analytic continuation a sufficiently
accurate initial guess for the iterative solution of
Eq.~\eqref{eq:gf3bI} on the real axis. This problem of the Matsubara
formulation is not related to the well known difficulty in performing
analytic continuation to the real axis.

All equations of motion are almost unchanged when we go over to
Matsubara frequency $i\omega_n$, {\it e.$\,$g.} Eq.~\eqref{eq:exactccgf}
becomes
\begin{equation}
\gl c_{kn};c^+_{qn} \gr_{i\omega_n} = \frac{\delta_{kq}}{i\omega_n - \varepsilon_q} + \frac{V_kV_q}{(i\omega_n - \varepsilon_k)(i\omega_n - \varepsilon_q)}\gl X^{0n};X^{n0}\gr_{i\omega_n}\,.
\label{eq:matsuexactccgf}\end{equation}
Correlations have to be calculated as
\begin{equation}\begin{split}
\langle c^+_{qn} c_{kn} \rangle = T\sum_{i\omega_n}\gl c_{kn}; c^+_{qn}\gr_{i\omega_n }e^{i\omega_n 0^+}
\label{eq:matsucorr}\end{split}\end{equation}
which replaces Eq.~\eqref{eq:correlation} for that purpose. 
In order to simplify the equations, we employ the analog of Eq.~\eqref{eq:identity}, namely
\begin{equation}
\frac{V_k^2}{(i\omega_n - \varepsilon_k)(i\omega'_n - \varepsilon_k)}=\frac{1}{i\omega'_n - i\omega_n}\biggl\{\frac{V_k^2}{i\omega_n - \varepsilon_k}-\frac{V_k^2}{i\omega'_n - \varepsilon_k}\biggr\} \,,
\end{equation}
and we identify occurrences of the hybridization function
\begin{equation}
\Delta(i\omega_n)=\sum_k\frac{V_k^2}{i\omega_n - \varepsilon_k}\,.
\label{eq:matsudelta}\end{equation}
This leads to the system of equations
\begin{equation}
F_m(i\omega_n) = \frac{1-n_f+\frac{n_f}{N}   +S_1(i\omega_n)}{ i\omega_n -\varepsilon_f- \Delta(i\omega_n)\bigl(1+S_1(i\omega_n)\bigr)+S_2(i\omega_n)} 
\label{eq:gf3dI}\end{equation}
with 
\begin{align}
S_1(i\omega_n)&=T \sum_{\substack{l\neq m\\i\omega'_n}}\frac{\Delta(i\omega'_n)-\Delta(i\omega_n)}{i\omega'_n - i\omega_n}F_l(i\omega'_n)e^{i\omega'_n 0^+}\label{eq:matsu_sum1}\\
S_2(i\omega_n)&=T \sum_{\substack{l\neq m\\i\omega'_n}}\frac{\Delta(i\omega'_n)-\Delta(i\omega_n)}{i\omega'_n - i\omega_n} \bigl\{ 1 + \Delta(i\omega'_n) F_l(i\omega'_n) \bigr\} e^{i\omega'_n 0^+}\label{eq:matsu_sum2}\\
n_f&=N T \sum_{i\omega'_n}F_m(i\omega'_n)e^{i\omega'_n 0^+}\label{eq:matsu_nn}\,.
\end{align}
With the replacement
\begin{equation}
T \sum_{i\omega'_n} K(i\omega'_n)e^{i\omega'_n 0^+} \quad\longrightarrow\quad -\frac{1}{\pi}\int d\omega'f(\omega')\imag  K(\omega')
\end{equation}
we can easily recover Eqs.~\eqref{eq:int1final} and
\eqref{eq:int2final} from \eqref{eq:matsu_sum1} and
\eqref{eq:matsu_sum2}.  It is important to note that good convergence
of the self-consistent solution of the system of equations depends
crucially on the proper treatment of the slowly decaying high
frequency tails of the addends of
Eqs.~\eqref{eq:matsu_sum1}-\eqref{eq:matsu_nn}.  A high frequency
expansion of these addends was performed to determine the coefficients
of the terms proportional to $1/i\omega_n$ and $1/(i\omega_n)^2$. These terms
were subtracted from the sums, and their value was determined
analytically.

\subsection{Hubbard model}

We now proceed to investigate the usefulness of the impurity solver
detailed above in its application in the DMFT context. Our application
of the method to three lattice models is an exploratory study
concentrating on a small number of important properties only. It is
not the intention of this article to go into detail for each of the
three models.  We first investigate the Hubbard model in order to
study the quasiparticle scaling of the Hubbard band with degeneracy
$N$.

We consider the Hubbard Hamiltonian
\begin{equation}
H=-\sum_{ij\sigma} t_{ij} c^+_{i\sigma}c_{j\sigma} -\mu \sum_{i\sigma} c^+_{i\sigma}c_{i\sigma} +\frac{U}{2} \sum_{\substack{i\sigma\sigma'\\\sigma\neq\sigma'}}c^+_{i\sigma}c_{i\sigma}c^+_{i\sigma'}c_{i\sigma'}
\label{eq:hubbard}\end{equation}
where the spin and orbital index $\sigma$ runs from 1 to $N$. For this
model, we have to solve the AIM with the self-consistency condition
\begin{equation}
\Delta(i\omega_n)=t^2G_{oo\,\sigma}(i\omega_n)\,.
\label{eq:selfcon}\end{equation}
For the derivation of this equation, see App. \ref{app:hubb}.

\subsection{Anderson lattice}

We study the application of the $U=\infty$ impurity solver to the Anderson lattice in order to learn how this new approach compares to the straightforward decoupling of the equations of motion for the periodic Anderson model~\cite{costi:86}.
We consider the periodic Anderson Hamiltonian 
\begin{equation}\begin{split}
H_{\text{PAM}} =& -\sum_{ij\sigma} t^c_{ij} c^+_{i\sigma}c_{j\sigma} -\sum_{ij\sigma} t^f_{ij} f^+_{i\sigma}f_{j\sigma} +\varepsilon_c \sum_{i\sigma} c^+_{i\sigma}c_{i\sigma} +\varepsilon_f \sum_{i\sigma} f^+_{i\sigma}f_{i\sigma} \\&+\sum_{i\sigma}\bigl(V_{i\sigma}c^+_{i\sigma}f_{i\sigma} +V^*_{i\sigma}f^+_{i\sigma}c_{i\sigma}\bigr)+\frac{U}{2} \sum_{\substack{i\sigma\sigma'\\\sigma\neq\sigma'}}f^+_{i\sigma}f_{i\sigma}f^+_{i\sigma'}f_{i\sigma'}
\end{split}\end{equation}
In this case, the self consistency condition for the $f$ electron
Greens function is
\begin{equation}
G_f^{\text{local}}(i\omega_n)
= \int d\varepsilon \,\rho_0(\varepsilon) \biggl(i\omega_n+\mu-\varepsilon_f^0-\Sigma_f(i\omega_n) -\frac{V(\varepsilon)^2}{i\omega_n+\mu-\varepsilon}\biggr)^{-1}\\
\end{equation}
Here, the self energy is determined from the equation 
\begin{equation}
\mathcal{G}_0^{-1}(i\omega_n)=G_f^{-1}(i\omega_n)+\Sigma(i\omega_n)
\label{eq:sigma}\end{equation}
and the Weiss function $\mathcal{G}_0(i\omega_n)$ is related to the
hybridization function $\Delta(i\omega_n)$ by
\begin{equation}
\mathcal{G}_0^{-1}(i\omega_n)=i\omega_n+\mu-\varepsilon_f^0-\Delta(i\omega_n)
\label{eq:weissfunction}\end{equation} 
The derivation of these equations is contained in App.~\ref{app:and}.

\subsection{$pd$ model}

In order to study the Mott transition with the $U=\infty$ impurity solver
described above, we consider the Hamiltonian~\cite{kotliar:93}
\begin{equation}\begin{split}
H=-\sum_{i j \sigma} V_{ij} \Bigl[d^+_{i\sigma}p_{j\sigma}+p^+_{j\sigma}d_{i\sigma}\Bigr] +\varepsilon_p\sum_{j \sigma}p^+_{j\sigma}p_{j\sigma} +\varepsilon_d\sum_{i \sigma}d^+_{i\sigma}d_{i\sigma} +U_d\sum_i d^+_{i\uparrow}d_{i\uparrow}d^+_{i\downarrow}d_{i\downarrow}\,.
\label{eq:Hmott}\end{split}\end{equation}
This Hamiltonian, which we call $pd$ model here, is similar to the
Anderson lattice Hamiltonian if the conduction electron dispersion is
taken to be a constant $\varepsilon_k=\varepsilon_p$ and if the $k$
dependence of the hybridization $V_k$ is retained. This changes the
local conduction electron Greens function:
\begin{equation}\begin{split}
G_d^{\text{local}}(i\omega_n) &= 
 \int d\varepsilon \,\rho_{pd}(\varepsilon) \biggl(i\omega_n+\mu-\varepsilon_d-\Sigma_d(i\omega_n) -\frac{\varepsilon^2}{i\omega_n+\mu-\varepsilon_p}\biggr)^{-1}\\
&=\zeta_p \int d\varepsilon \,\rho_{pd}(\varepsilon)\biggl(\zeta_p \zeta_d-\varepsilon^2 \biggr)^{-1}\\
G_p^{\text{local}}(i\omega_n) &= 
\int d\varepsilon \,\rho_{pd}(\varepsilon) \biggl( i\omega_n+\mu-\varepsilon_p-\frac{\varepsilon^2}{i\omega_n+\mu-\varepsilon_d-\Sigma_d(i\omega_n)}\biggr)^{-1}\\
&=\zeta_d \int d\varepsilon \,\rho_{pd}(\varepsilon)\biggl(\zeta_p \zeta_d-\varepsilon^2 \biggr)^{-1} 
\label{eq:Gmott}\end{split}\end{equation}
 with the abbreviations $\zeta_p=i\omega_n+\mu-\varepsilon_p$ and
 $\zeta_d=i\omega_n+\mu-\varepsilon_d-\Sigma_d(i\omega_n)$. Here, $\rho_{pd}(\varepsilon)$ stands for the density
 of states associated with the hybridization $V_k$.  Noting that
\begin{equation}\begin{split}
\int d\varepsilon \,\frac{\rho_{pd}(\varepsilon)}{x^2-\varepsilon^2}&=\frac{1}{2x}\biggl[\int d\varepsilon \,\frac{\rho_{pd}(\varepsilon)}{x-\varepsilon}+\int d\varepsilon \,\frac{\rho_{pd}(\varepsilon)}{x+\varepsilon}\biggr]=\frac{1}{2x}\bigl(\tilde{D}(x)-\tilde{D}(-x)\bigr)\\
&=\frac{\tilde{D}(x)}{x}\qquad \text{for symmetric }\rho_{pd}(\varepsilon) 
\end{split}\end{equation}
we can write $G_p^{\text{local}}(i\omega_n)$ and $G_d^{\text{local}}(i\omega_n)$
as Hilbert transforms
\begin{equation}\begin{split}
G_p^{\text{local}}(i\omega_n)&=\sqrt{\frac{\zeta_d}{\zeta_p}}\tilde{D}\biggl(\sqrt{\zeta_p \zeta_d}\biggr)\\
G_d^{\text{local}}(i\omega_n)&=\sqrt{\frac{\zeta_p}{\zeta_d}}\tilde{D}\biggl(\sqrt{\zeta_p \zeta_d}\biggr)
\label{eq:GmottI}\end{split}\end{equation}
We use a semicircular form for $\rho_{pd}(\varepsilon)$:
\begin{equation}
\rho_{pd}(\varepsilon) = \frac{1}{2\pi t_{pd}^2} \sqrt{4t_{pd}^2-\varepsilon^2}
\end{equation}
where $t_{pd}$ is the strength of the hybridization between $p$ and
$d$ levels.

It is worth pointing out, that this method reproduces an important
aspect of the exact solution of the DMFT equations within the context
of the $pd$ model. Namely, it produces a first order phase transition
between a metallic and an insulating phase, which is manifest by the
existence of two DMFT solutions for the same range of parameters.

\section{Results}

\subsection{Hubbard model}

\begin{figure}
\begin{center}
\includegraphics[height=\textwidth,angle=-90]{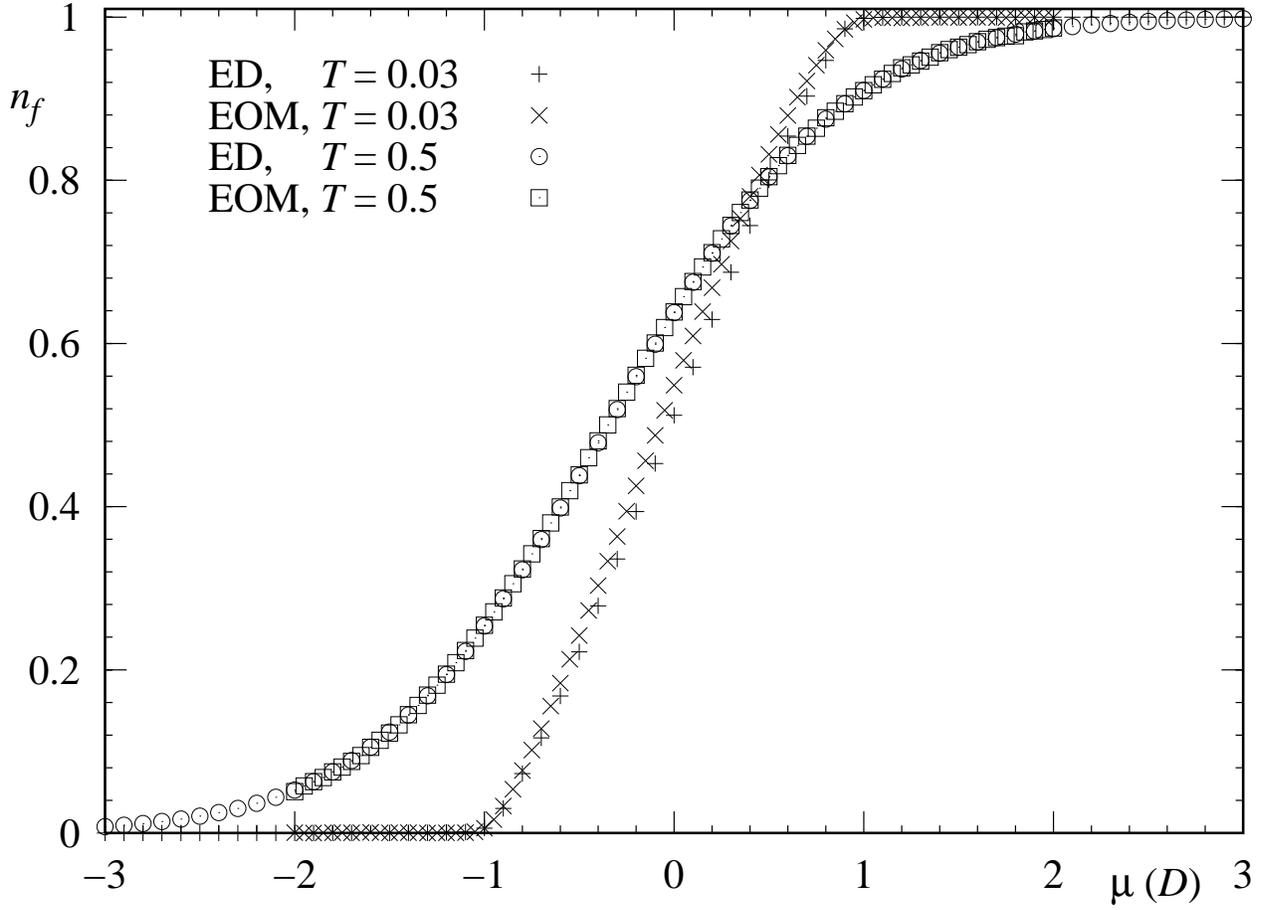}
\end{center}
\caption{
  Density of $f$ electrons as a function of the chemical potential
  $\mu=-\varepsilon_f$ for exact diagonalization and the equation of motion method
  in comparison. The energy unit is the half band width $D$. For the
  higher temperature $T=0.5$, the two methods agree extremely well,
  while for the lower temperature $T=0.03$, the exact diagonalization
  gives slightly lower densities at the same chemical potential
  $\mu=-\varepsilon_f$. Exact diagonalization was performed with 6 sites, and the
  Hubbard model was solved in the DMFT approximation.}
\label{fig:ed_eom_compare_nn}\end{figure}

First of all we test the performance of our impurity solver by
comparing it with the results of exact diagonalization (ED). For this
purpose, we employ the code published accompanying the review of the
DMFT method in Ref.~\cite{kotliar:96}, modified to $U=\infty$. The Hubbard
model is solved in the DMFT approximation. The self-consistency
condition for the Hubbard model is realized by minimizing the function
$f(\varepsilon_k,V_k)=\sum_n|t^2 G(i\omega_n)-\sum_k V_k^2/(i\omega_n-\varepsilon_k)|$ with respect to the
parameters $\varepsilon_k$ and $V_k$. Here, the exact diagonalization has been
performed with $N_s=6$ sites which are divided into 1 site for the
impurity and 5 sites for the bath.  Thus, the hybridization function
$\Delta(i\omega_n)$ is represented with 5 poles. This leads to a finite number
of poles instead of a smooth function in the spectral function as
well. Fig.~\ref{fig:ed_eom_compare_nn} shows the comparison of the
densities of f electrons as a function of the impurity position $\varepsilon_f$
(which is related to the chemical potential by $\mu= - \varepsilon_f$). The
comparison shows that at high temperature $T=0.5$, the results of
exact diagonalization and EOM are virtually indistinguishable while
for a lower temperature $T=0.03$, the densities differ slightly for
impurity positions between -1 and 1.

\begin{figure}
\begin{center}
\includegraphics[height=\textwidth,angle=-90]{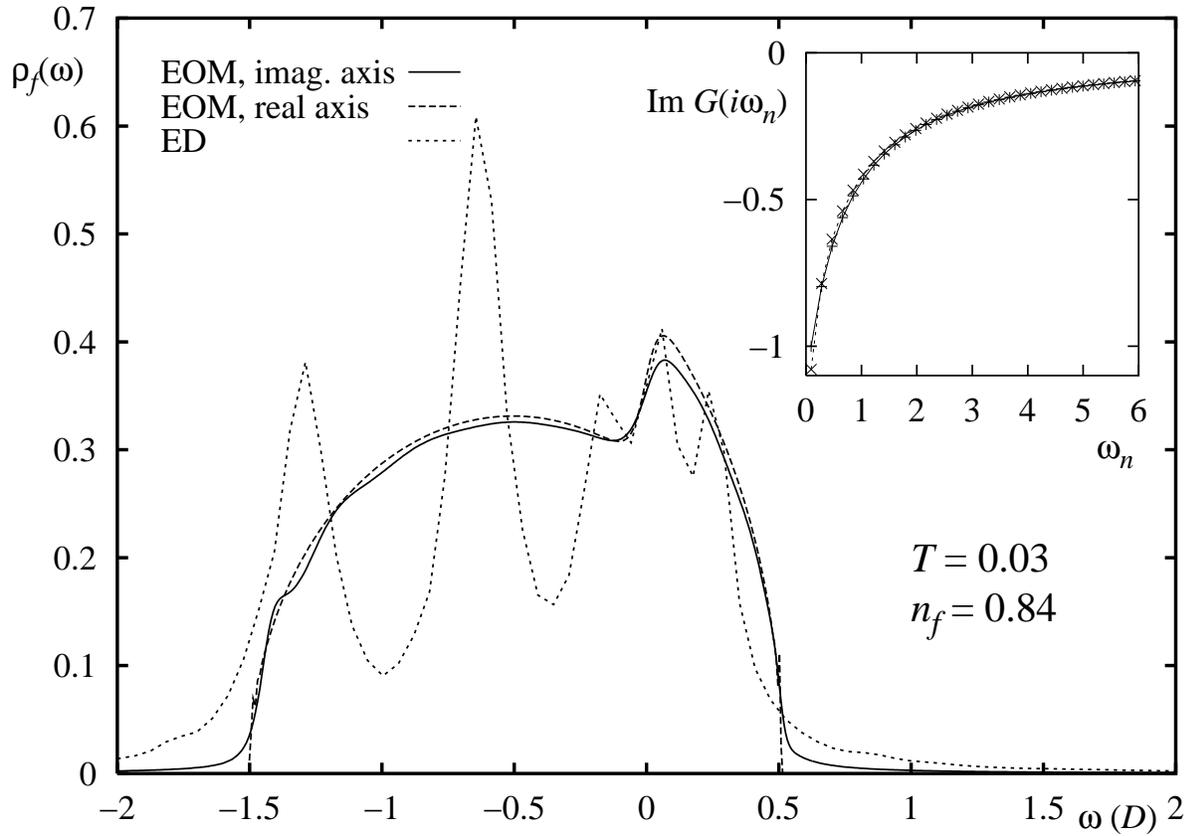}
\end{center}
\caption{
  Spectral functions (main figure) and imaginary part of the Greens
  function (inset) from exact diagonalization and the equation of
  motion method in comparison. The temperature is $T=0.03$, the
  density of $f$ electrons is $n_f =0.84$ for both methods. The two
  methods compare well, considering that the exact diagonalization
  with 5 bath sites has only limited resolution on the real axis.  }
\label{fig:ed_eom_compare_dos}\end{figure}

In Fig.~\ref{fig:ed_eom_compare_dos}, we compare the imaginary parts
of the Greens function for a density of $n_f=0.84$. The slight
differences in the $n_f$ versus $\mu$ curves of
Fig.~\ref{fig:ed_eom_compare_nn} mean that this density is achieved
for $\mu=0.6$ in the case of ED and for $\mu=0.53$ in the case of EOM. The
imaginary parts of the Greens function on the Matsubara axis shown in
the inset are very similar. Thus, the main figure shows the more
demanding comparison of the densities of state. The continuous line
represents the DOS from the EOM method gained by analytic continuation
in the Pad{\'e} approximation, while the long dashes stand for the EOM
result on the real axis.  The dashed curve with the five poles is the
result of DMFT on the basis of exact diagonalization. The figure shows
that the distribution of spectral weight between the Kondo peak around
the Fermi level at $\omega=0$ and the Hubbard band is similar in both
methods, but the EOM method leads to a better overall shape of the
spectral function.  We conclude that the EOM method results compare
well with ED, giving us confidence that it is a useful
approximation. Even for this low number of $N_s=6$ sites, the exact
diagonalization requires an order of magnitude more CPU time than the
EOM method.

\begin{figure}
\begin{center}
\includegraphics[height=0.8\textwidth,angle=-90]{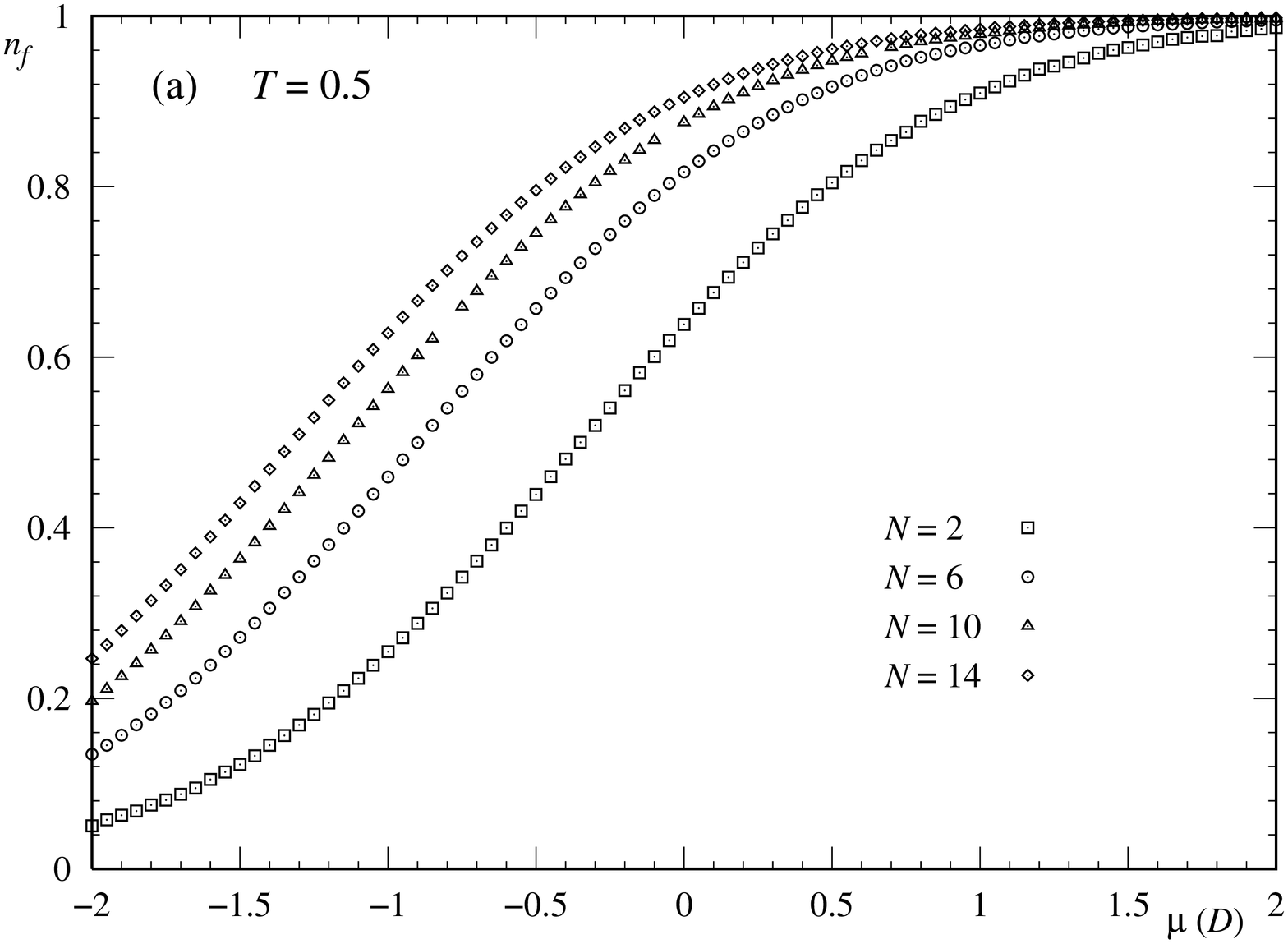}
\vspace{0.7cm}\\
\includegraphics[height=0.8\textwidth,angle=-90]{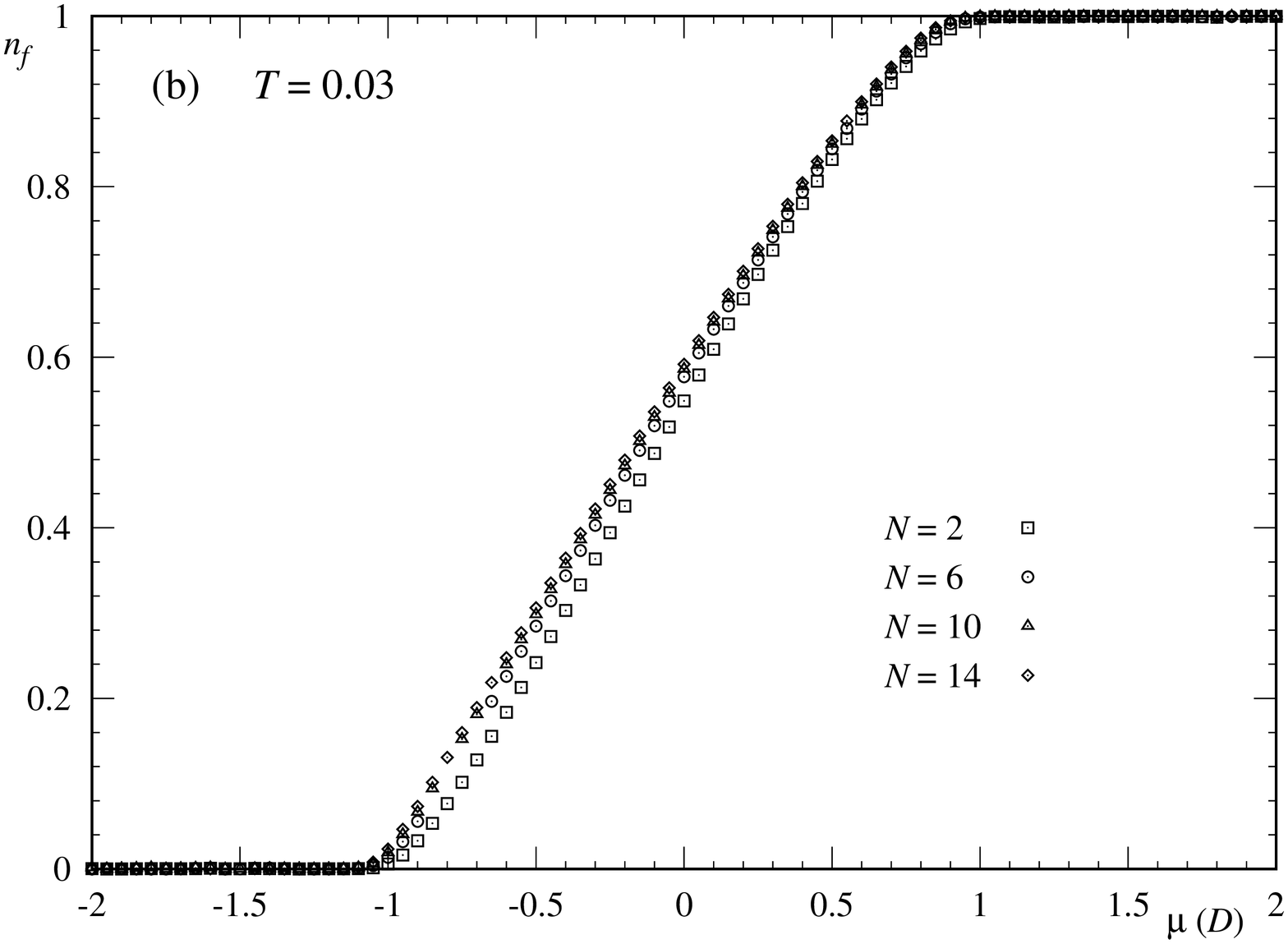}
\end{center}
\caption{
  Density $n_f$ of $f$ electrons as a function of the chemical
  potenital $\mu$ for the infinite $U$ Hubbard model. Energy is measured
  in units of half band width $D$. (a) At high temperature $T=0.5$,
  $n_f(\varepsilon_f)$ differs for different values of the degeneracy $N$. (b)
  At low temperature $T=0.03$, the $n_f(\varepsilon_f)$ for different $N$ nearly
  coincide.  }
\label{fig:hubbard_nversusmu}\end{figure}

Fig.~\ref{fig:hubbard_nversusmu} shows the carrier density as a
function of the impurity position $\varepsilon_f$. The impurity position
corresponds to the chemical potential, only with opposite sign
$\mu=-\varepsilon_f$. Due to the infinite interaction, the maximum filling is 1
electron per site. In other words, the upper Hubbard band that could
hold a second electron at finite $U$ has been pushed to infinite
energy.  While at low temperature $T=0.03$ the $n_f$ versus $\mu$ curves
at different degeneracies $N=2$ to $N=14$ nearly coincide (see
Fig.~\ref{fig:hubbard_nversusmu}~(b)), they differ considerably at
high temperature $T=0.5$ (see
Fig.~\ref{fig:hubbard_nversusmu}~(a)). %(Grund?)

\begin{figure}
\begin{center}
\includegraphics[height=0.8\textwidth,angle=-90]{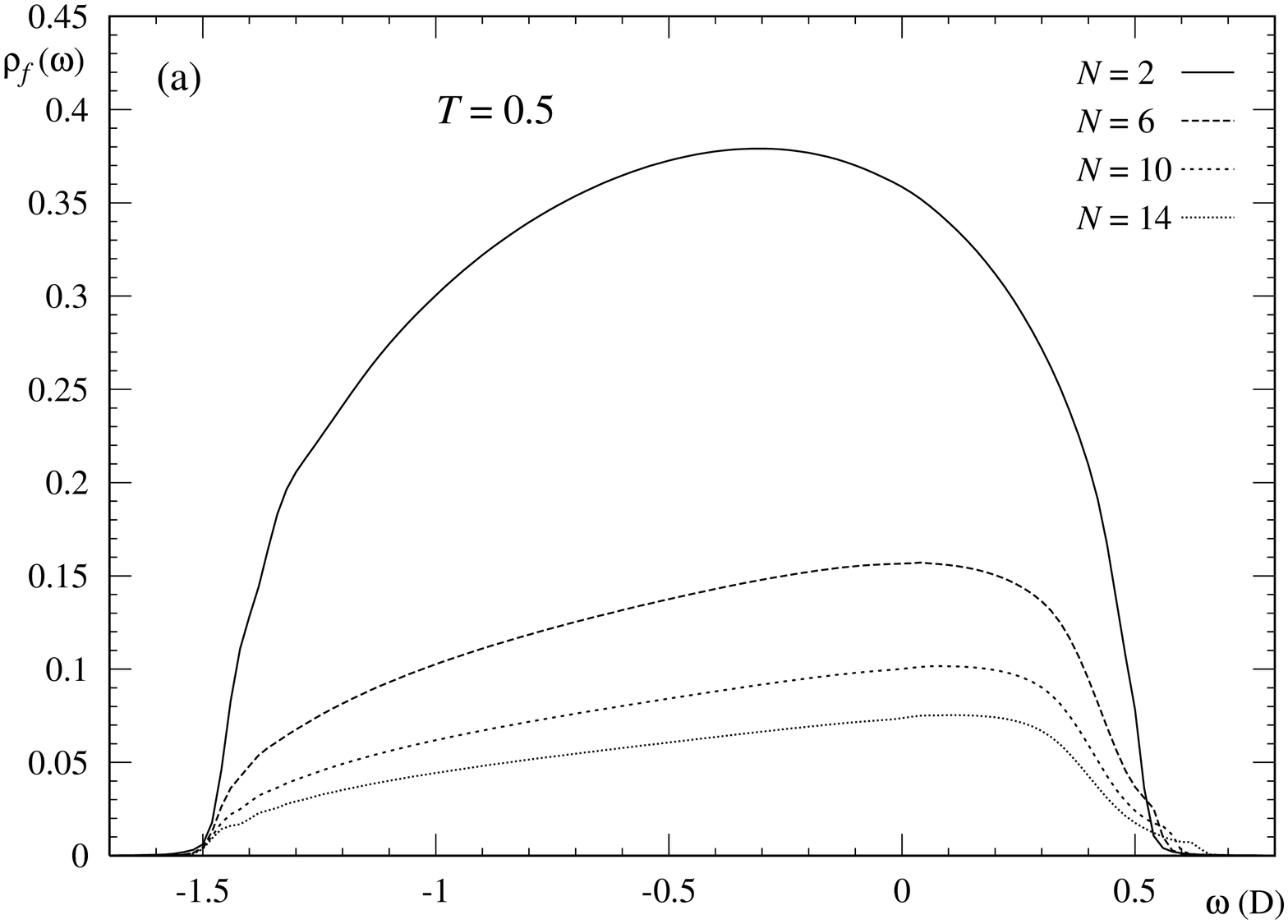}
\vspace{0.7cm}\\
\includegraphics[height=0.8\textwidth,angle=-90]{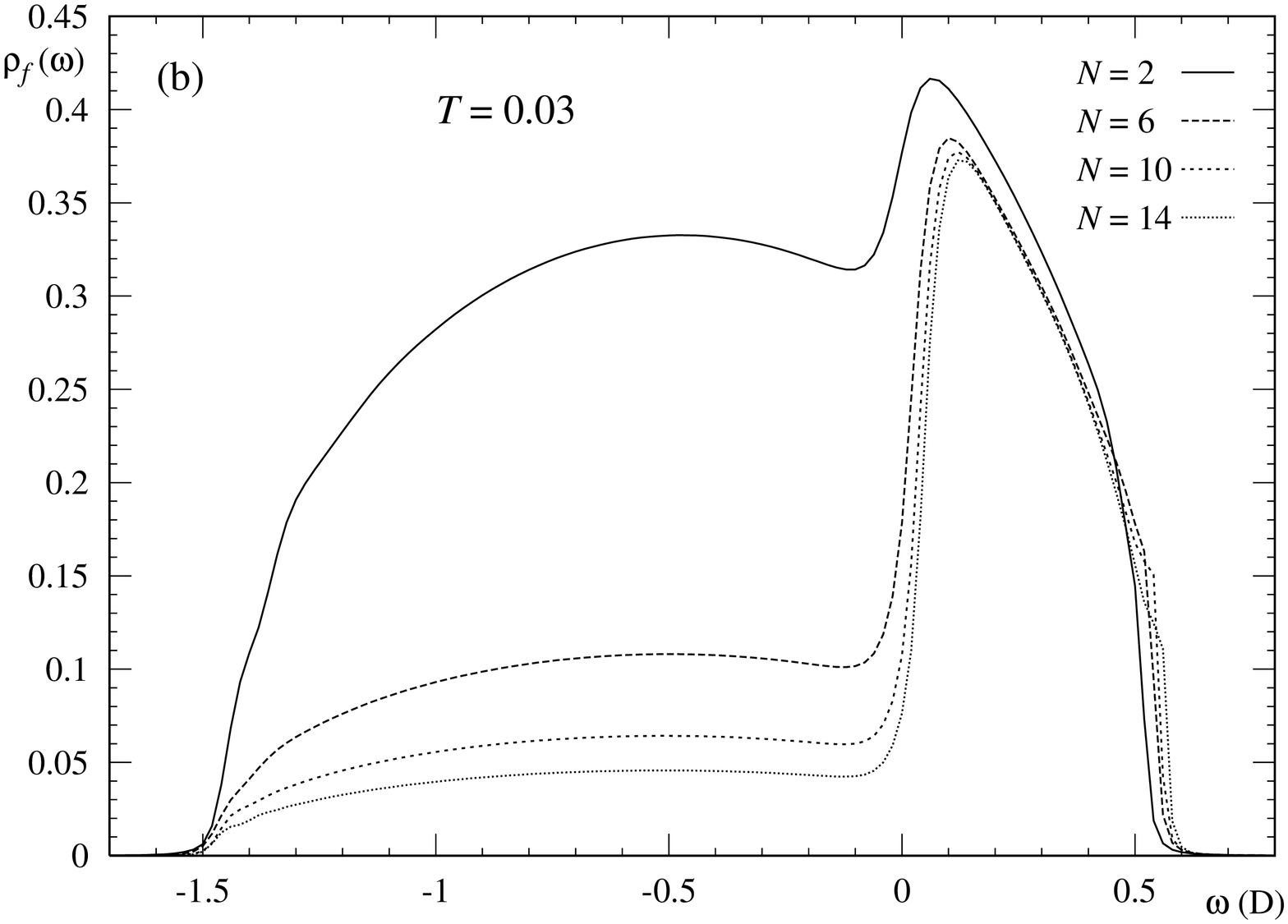}
\end{center}
\caption{
  Density of states $\rho_f(\omega)$ of $f$ electrons for the infinite $U$
  Hubbard model. The energy unit is the half band width $D$. (a) At
  high temperature $T=0.5$, there is no quasiparticle resonance at
  $\omega=0$. (b) At low temperature $T=0.03$, the quasiparticle resonance
  at $\omega=0$ is clearly developed. The weight of the Hubbard band is
  proportional to $1/N$.  }
\label{fig:hubbard_em0p5_dos}\end{figure}

In Fig.~\ref{fig:hubbard_em0p5_dos} we show examples of the spectral
function for degeneracies between $N=2$ and $N=14$ for high and low
temperature. While at $T=0.5$ the spectral function is nearly
unstructured, at $T=0.003$ a broad Hubbard band and a quasiparticle
resonance at zero frequency $\omega=0$ can be distinguished. The weight of
the Hubbard band diminishes as $1/N$ as the degeneracy $N$ increases
while the intensity of the Kondo peaks remains nearly constant.
Note that the spectral functions in Fig.~\ref{fig:hubbard_em0p5_dos}
resulting from the DMFT self-consistency contain no spurious side bands
as those calculated by directly decoupling the equations of motion
produced by the Hubbard Hamiltonian~\cite{gros:94}. In our
calculation, the imaginary part of the Greens function outside the
Hubbard band and resonance is exactly zero.

%[comp. with ED?]
%[uvs. \mathmicro curves for PAM?]
\subsection{Anderson lattice}

\begin{figure}
\begin{center}
\includegraphics[height=0.8\textwidth,angle=-90]{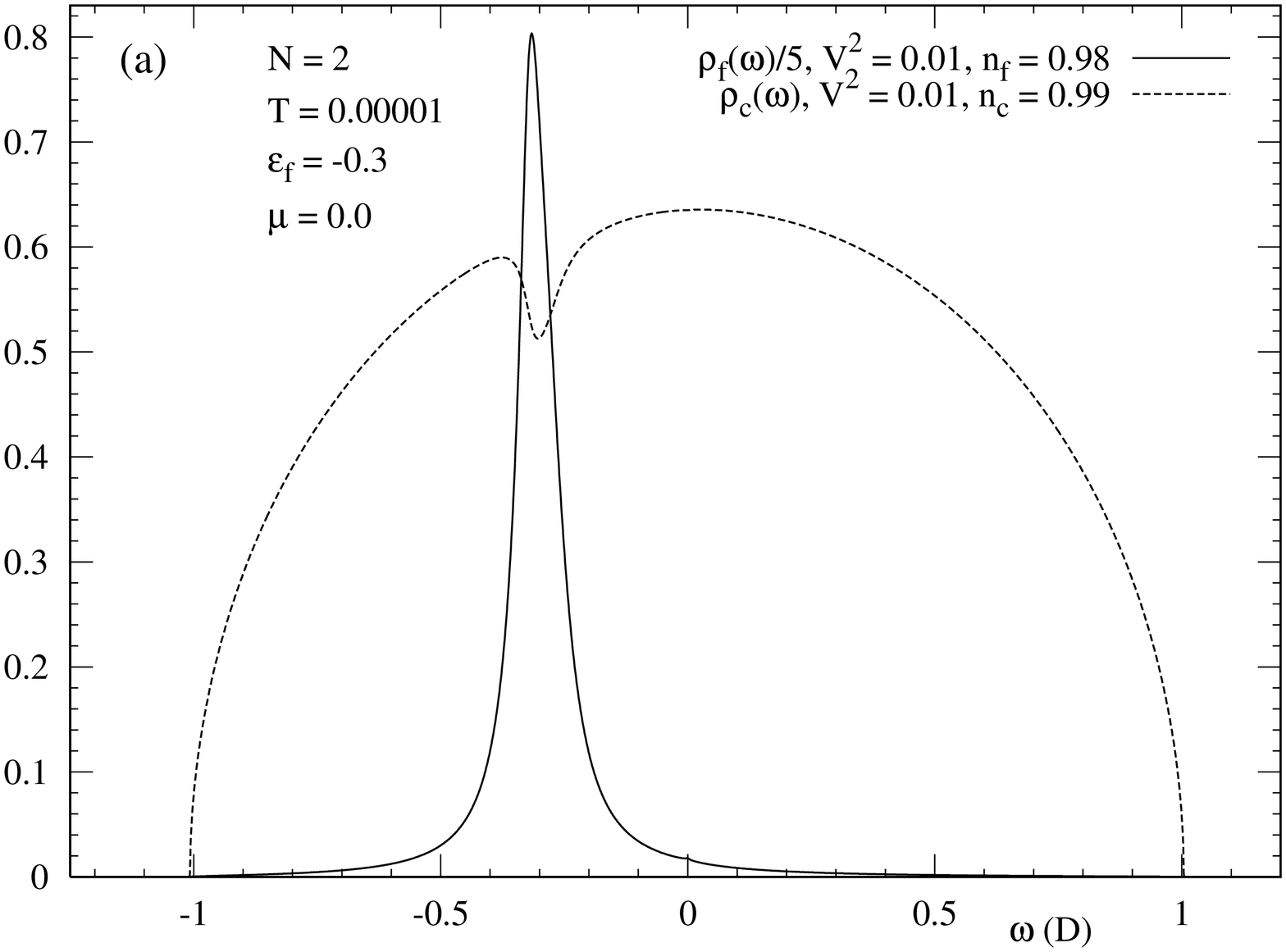}
\vspace{0.7cm}\\
\includegraphics[height=0.8\textwidth,angle=-90]{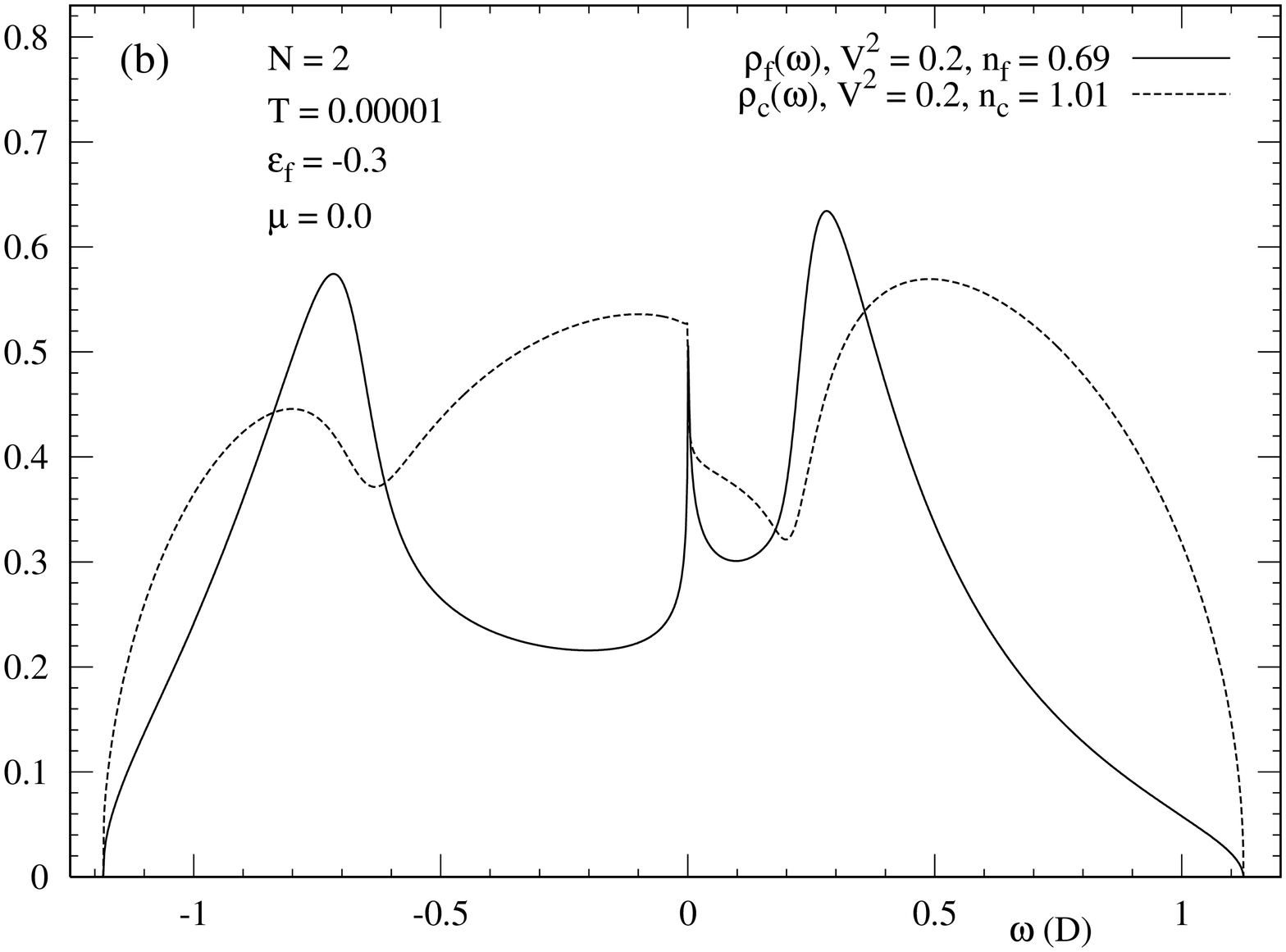}
\end{center}
\caption{
  Densities of states $\rho_f(\omega)$ and $\rho_c(\omega)$ of $f$ and conduction
  electrons for the infinite $U$ periodic Anderson model. (a) At a low
  hybridization $V^2=0.01$, the $f$ electron Greens function is maily
  a peak at the impurity position; there is no quasiparticle resonance
  at $\omega=0$. (b) At a high hybridization $V^2=0.2$, the hubbard band of
  the $f$ electron Greens function is split into two peaks, and the
  quasiparticle resonance at $\omega=0$ is clearly developed. }
\label{fig:andT0.00001_Vcompare}\end{figure}

Fig.~\ref{fig:andT0.00001_Vcompare} shows examples for the conduction
electron and the strongly correlated $f$ electron spectral functions
(dashed and full lines, respectively). In
Fig.~\ref{fig:andT0.00001_Vcompare}~(a), the hybridization between the
two bands is small ($V^2=0.01$) while in
Fig.~\ref{fig:andT0.00001_Vcompare}~(b) it is rather large
($V^2=0.2$). Correspondingly, the conduction electron DOS shows only a
small dip at the position of the f band for a low value of the
hybridization.  Interestingly, we find a Kondo resonance at the Fermi
level in the $f$ electron DOS. This resonance was absent in the
decoupling approach to the periodic Anderson model of
Ref.~\cite{costi:86}.
 
\subsection{$pd$ model}

\begin{figure}
\begin{center}
\includegraphics[height=0.8\textwidth,angle=-90]{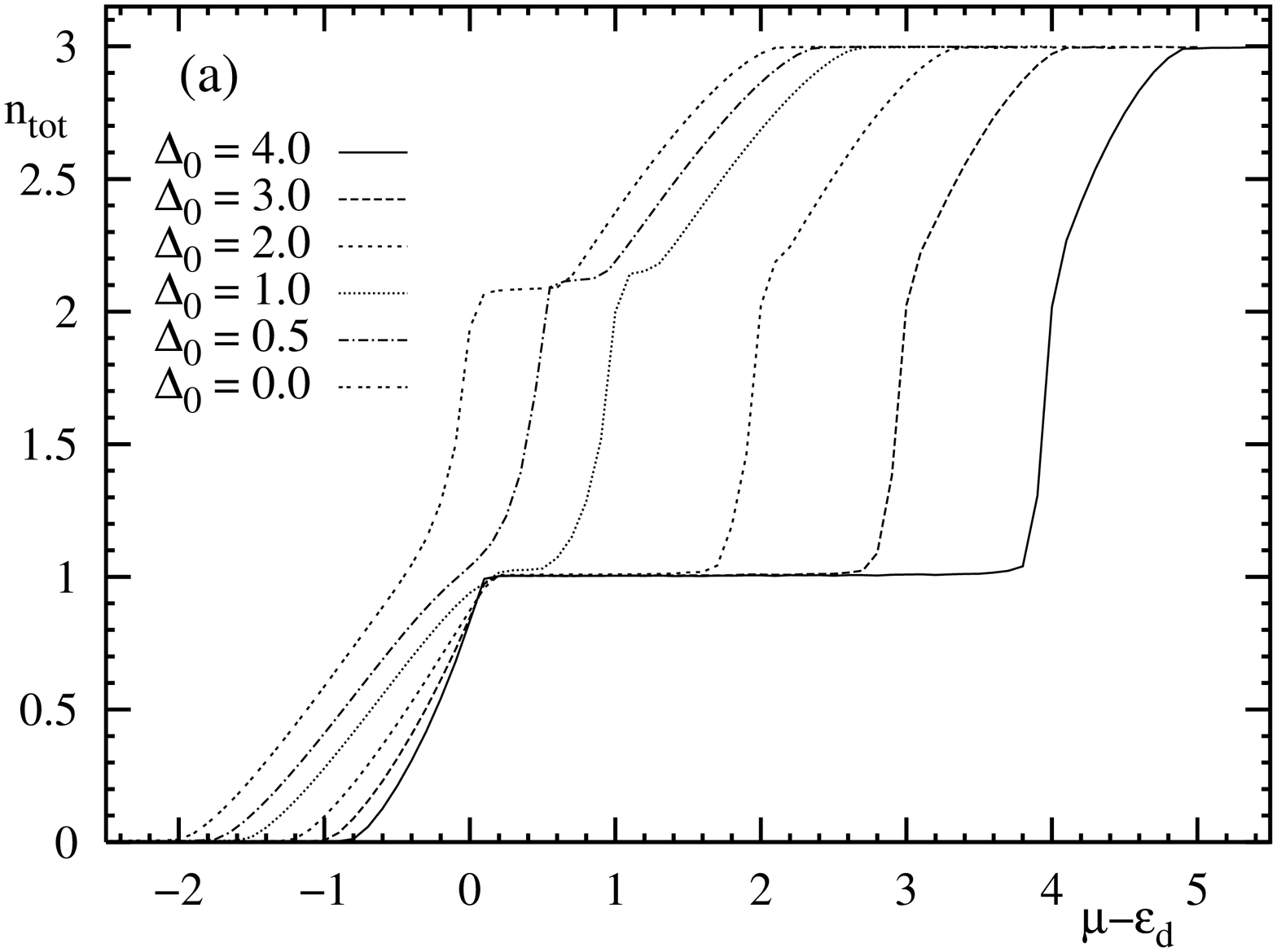}
\\
\includegraphics[height=0.8\textwidth,angle=-90]{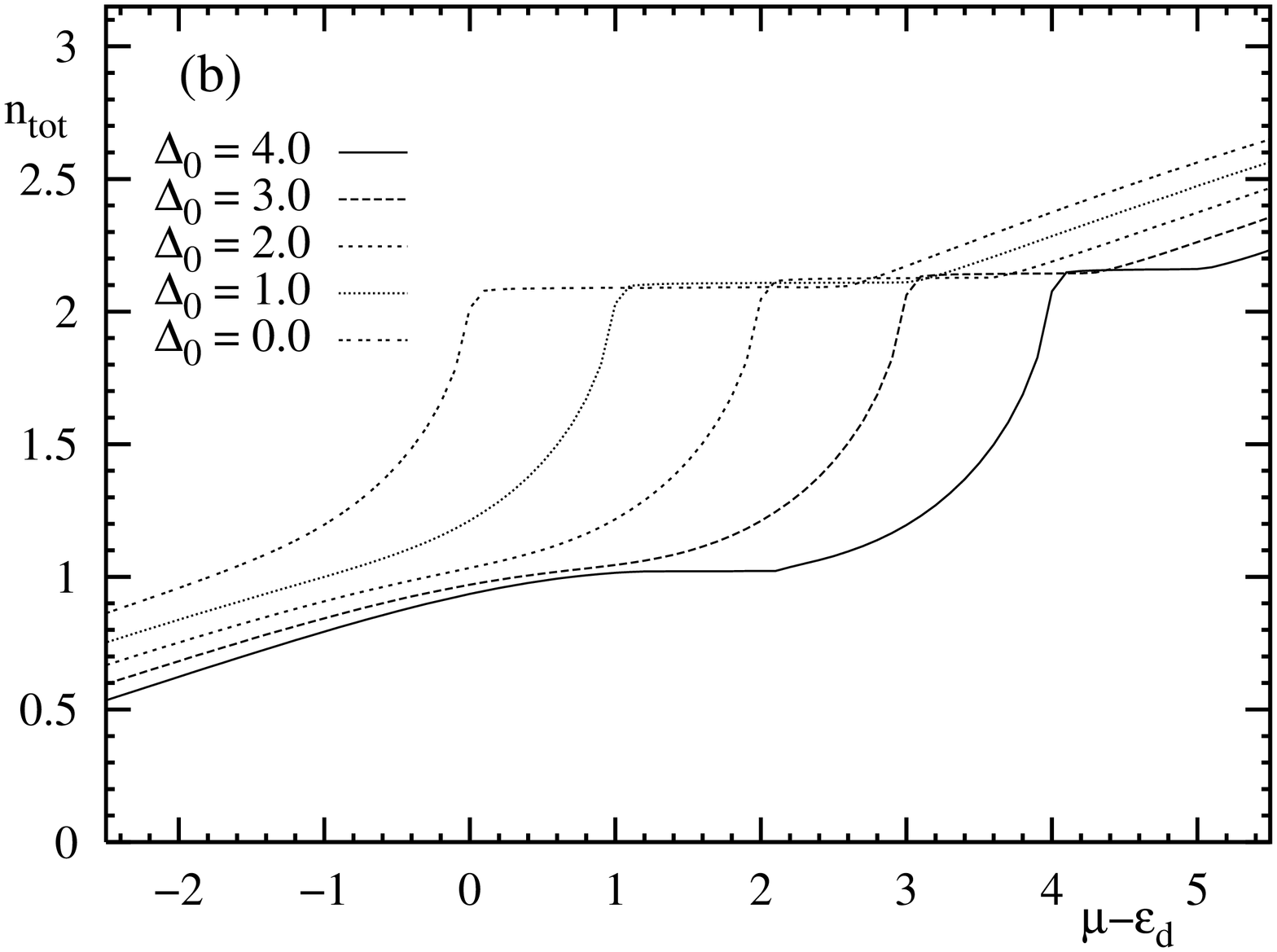}
\end{center}
\caption{
  Total density of electrons as a function of chemical potential for
  the $pd$ model Hamiltonian \eqref{eq:Hmott}. The hybridization
  strengths were (a) $t_{pd} = 1$ and (b) $t_{pd} = 4$. The plateaus
  at densities $n_{total}=1$ and $n_{total}=2$ correspond to the
  insulating phases.  }
\label{fig:mott_density}\end{figure}

We investigate the $pd$ model Hamiltonian Eq.~\eqref{eq:Hmott} as a
function of the separation $\Delta_0=\varepsilon_p - \varepsilon_d$ and of the hybridization
strength $t_{pd}$ between the two bands. From the analysis in
Ref.~\cite{kotliar:93} of the finite $U$ version of this model, we expect
a metal insulator transition to occur at a fixed density $n_{tot}=1$
if we vary the level separation $\Delta_0$ at a given
$t_{pd}$. Fig.~\ref{fig:mott_density}~(a) a shows the result of this
calculation at a fixed $t_{pd}=1$. The temperature was taken to be
$T=0.01$. For level separations $\Delta_0=0$ and $\Delta_0=0.5$ , the density
$n_{tot}$ around $n_{tot}=1$ changes smoothly as a function of
chemical potential. But beginning with $\Delta_0=1.0$, a charge transfer
gap $g_1=\mu(n_{tot}=1^+) - \mu(n_{tot}=1^-)$ begins to open up. Thus, the
physics discussed in Ref.~\cite{kotliar:93} for finite values of $U$ can
be also found for $U=\infty$. The critical value at $t_{pd}=1$ is
$\Delta_0=1$. Note that the $\Delta_0=4$ $U=8$ result in Fig. 1 of
Ref.~\cite{kotliar:93} compares well with the $\Delta_0=4$ $U=\infty$ curve of this
work's Fig.~\ref{fig:mott_density}~(a).  If we increase the
hybridization strength to $t_{pd}=4$ (see
Fig.~\ref{fig:mott_density}~(b)), we find that the critical $\Delta_0$ for
the metal to charge transfer insulator increases to $\Delta_0 \approx 4$.  In
Fig.~\ref{fig:mott_density}~(a), we also note the transition at a
total density $n_{tot}=2$ from a metal at higher level separation
$\Delta_0$ to a band insulator with a gap $g_2=\mu(n_{tot}=2^+) -
\mu(n_{tot}=2^-)$ .  For the higher value of the hybridization strength
$t_{pd}$ , the system is a band insulator at $n_{tot}=2$ for all
studied level separations $\Delta_0$.

\begin{figure}
\begin{center}
\includegraphics[height=0.8\textwidth,angle=-90]{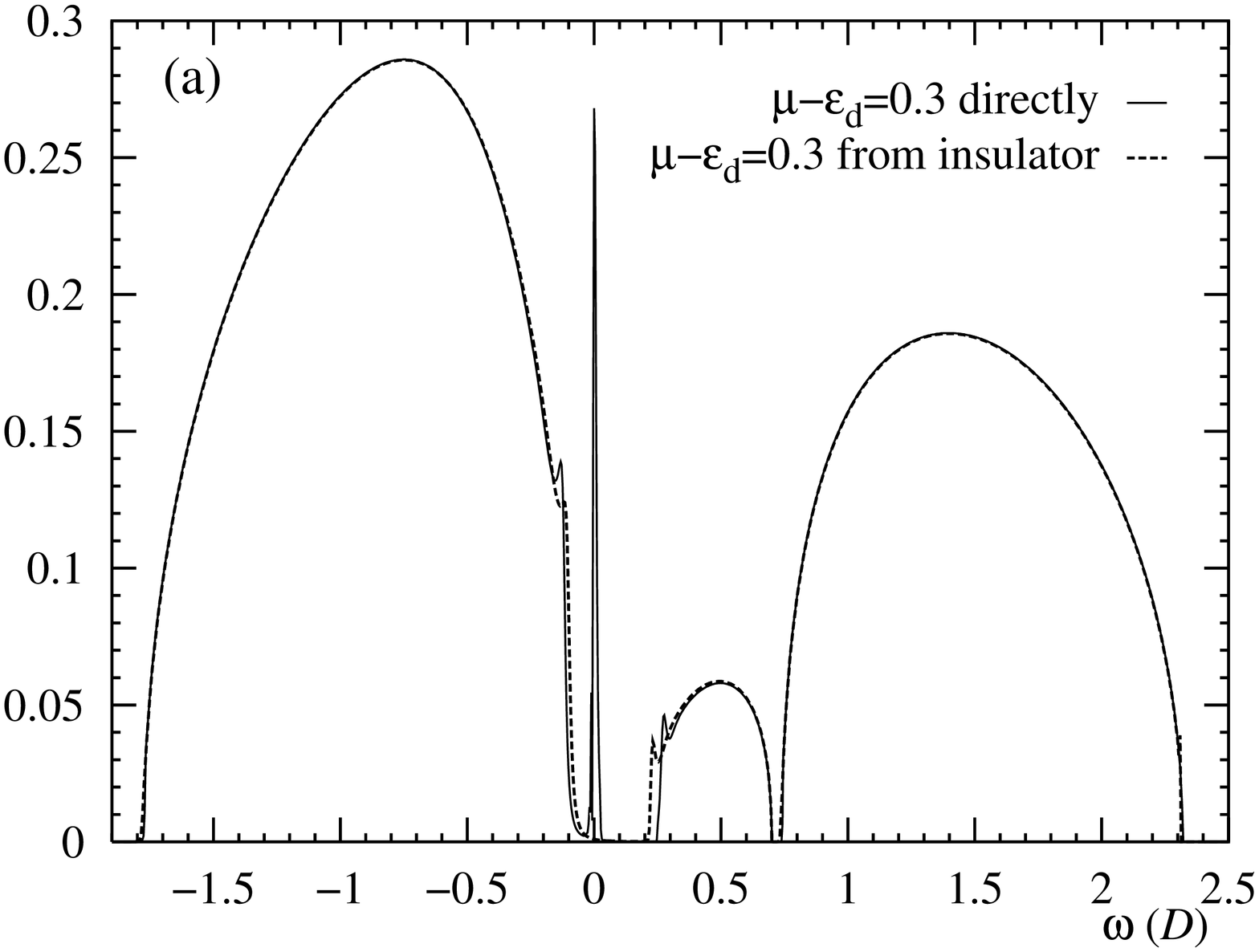}
\\
\includegraphics[height=0.8\textwidth,angle=-90]{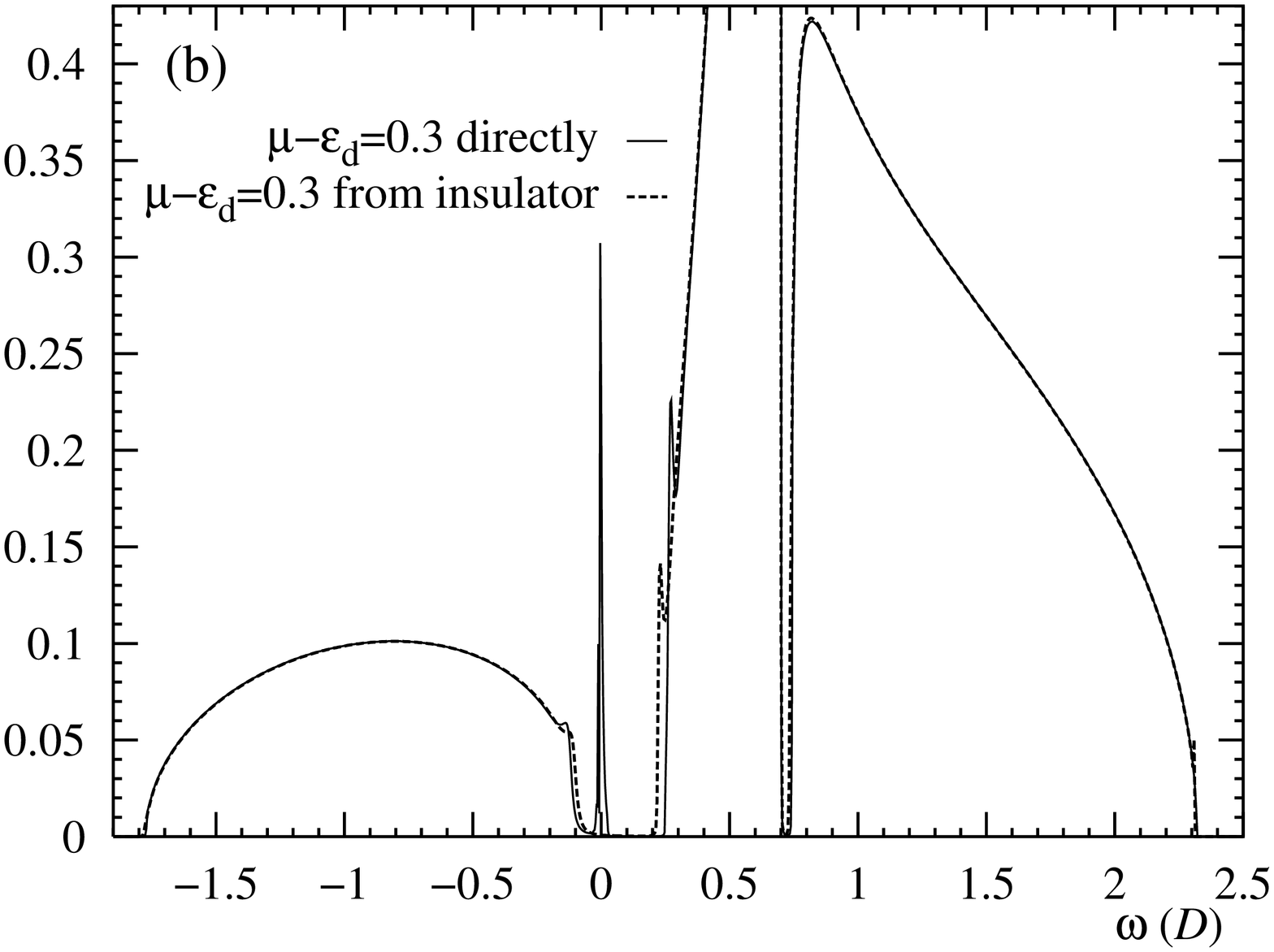}
\end{center}
\caption{
  Spectral function for the $pd$ model Hamiltonian \eqref{eq:Hmott}
  showing the coexistence of a metallic and an insulating phase. The
  hybridization strength was $t_{pd} = 1$, the $pd$ separation
  $\Delta_0=1$. (a) shows the correlated $d$ spectral function, (b) the
  uncorrelated $p$ spectral function. The full line corresponds to a
  metallic solution, the dashed line to an insulating solution. In (b)
  the sharp peak of the noninteracting DOS at $\varepsilon_p=0.5$ is not shown.
}
\label{fig:mott_coexistence}\end{figure}

An important question in the metal to insulator transition of
Fig.~\ref{fig:mott_density} concerns the existence of a coexistence
region. We can show that such a coexistence is indeed found with our
method. Fig.~\ref{fig:mott_coexistence} shows spectral functions for
$d$ and $p$ electrons at a hybridization strength $t_{pd}=1$, a
separation $\Delta_0=\varepsilon_p - \varepsilon_d=1$ and a chemical potential $\mu-\varepsilon_d=0.3$. The
calculation was performed for a temperature of $T=10^{-5}$, and care
was taken to resolve the sharp peak of the noninteracting Greens
function $G_p(\omega)$ at $\varepsilon_p=0.5$ with the help of a logarithmic grid.
The full line shows the converged result of a direct calculation at
$\mu-\varepsilon_d=0.3$. A quasiparticle peak at $\omega=0$ for both the correlated and
the uncorrelated electrons makes this a metallic solution. The dashed
line was obtained by starting the calculation in the insulating region
at $\mu-\varepsilon_d=0.5$ and lowering the chemical potential in steps of
$0.01$. At $\mu-\varepsilon_d=0.3$, the solution is still insulating as no
quasiparticle peak has formed.

\section{Summary}

A method to solve the Anderson impurity model with the help of
equations of motion and decoupling has been tested for its suitability
as an impurity solver in the framework of dynamical mean field theory.
The application to three lattice models in infinite dimensions and for
infinite interaction strength $U$ shows very encouraging results. In
the application to the Hubbard model, we see a correct quasiparticle
scaling of the Hubbard band with respect to the degeneracy. In the
periodic Anderson model, we find a Kondo resonance which is absent in
a direct decoupling of the equations of motion. This underlines the
usefulness of the approach chosen here: To use a decoupling scheme for
the solution of the Anderson impurity model which is then employed to
solve lattice models in the DFMT approximation. Interestingly, the
application of our approach to the $pd$ model yields a coexistence of
the insulating and metallic phases. The extension of the $U=\infty$
approach discussed here to finite values of the interaction strength
$U$ is possible and in preparation.  The numerical efficiency of the
method makes an application in an LDA+DMFT context feasible.

\section{Acknowledgments}
HOJ gratefully acknowledges support from the Deutsche
Forschungsgemeinschaft (DFG) through the Emmy Noether Programme. He
also wishes to thank for useful discussions with Theo Costi, Kristjan
Haule and Sarma Kancharla. GK is supported by NSF DMR-0096462.

\begin{appendix}

\section{DMFT self consistency condition for the Hubbard model}\label{app:hubb}

The partition function corresponding to the Hamiltonian of
Eq.~\eqref{eq:hubbard} is
\begin{equation}
Z=\int \prod_i  \mathcal{D}\bar{c}_{i\sigma}\mathcal{D}c_{i\sigma}  e^{-S}
\end{equation}
with the action
\begin{equation}\begin{split}
S=\int_0^\beta d\tau \sum_{i\sigma} \bar{c}_{i\sigma}(\tau)\frac{\partial}{\partial\tau}c_{i\sigma}(\tau) + \int_0^\beta d\tau \biggl[&-\sum_{ij\sigma} t_{ij} \bar{c}_{i\sigma}(\tau)c_{j\sigma}(\tau) -\mu \sum_{i\sigma} \bar{c}_{i\sigma}(\tau)c_{i\sigma}(\tau) \\&+\frac{U}{2} \sum_{\substack{i\sigma\sigma'\\\sigma\neq\sigma'}} \bar{c}_{i\sigma}(\tau)c_{i\sigma}(\tau)\bar{c}_{i\sigma'}(\tau)c_{i\sigma'}(\tau) \biggr]
\label{eq:action}\end{split}\end{equation} 
where the Fermion operators $c^+_{i\sigma}$, $c_{i\sigma}$ of the Hamiltonian
have been replaced by Grassmann variables $\bar{c}_{i\sigma}(\tau)$,
$c_{i\sigma}(\tau)$.

The cavity method now requires that we focus on one site $i=o$ and
separate the Hamiltonian \eqref{eq:hubbard} into three parts, one
relating to site $o$ only, one connecting this site to the lattice and
one for the lattice with site $o$ removed:
\begin{align}
H&=H_o + H_c + H^{(o)}\\
H_o&= -\mu \sum_{\sigma} c^+_{o\sigma}c_{o\sigma}+\frac{U}{2} \sum_{\substack{\sigma\sigma'\\\sigma\neq\sigma'}}c^+_{o\sigma}c_{o\sigma}c^+_{o\sigma'}c_{o\sigma'}\\
H_c&=-\sum_{i\sigma} \Bigl[ t_{io} c^+_{i\sigma}c_{o\sigma} +t_{oi} c^+_{o\sigma}c_{i\sigma}\Bigr]\\
H^{(o)}&= -\sum_{i\neq o\,j\neq o\,\sigma} t_{ij} c^+_{i\sigma}c_{j\sigma}-\mu \sum_{i\neq o\,\sigma} c^+_{i\sigma}c_{i\sigma} +\frac{U}{2} \sum_{\substack{i\neq o\,\sigma\sigma'\\\sigma\neq\sigma'}}c^+_{i\sigma}c_{i\sigma}c^+_{i\sigma'}c_{i\sigma'}
\end{align}
The three parts of the Hamiltonian correspond to the action $S_o$ of
site $o$, the action $\Delta S$ for the interaction between site $o$ and
the lattice, and the action $S^{(o)}$ of the lattice without site $o$:
\begin{align}
S_o=&\int_0^\beta d\tau \biggl[\sum_{\sigma} \bar{c}_{o\sigma}(\tau)\Bigl(\frac{\partial}{\partial\tau}-\mu\Bigr)c_{o\sigma}(\tau)  +\frac{U}{2} \sum_{\substack{\sigma\sigma'\\\sigma\neq\sigma'}} \bar{c}_{o\sigma}(\tau)c_{o\sigma}(\tau)\bar{c}_{o\sigma'}(\tau)c_{o\sigma'}(\tau) \biggr]\\
\Delta S=& -\int_0^\beta d\tau \biggl[ \sum_{i\sigma} t_{io}  \bar{c}_{i\sigma}(\tau)c_{o\sigma}(\tau) +t_{oi} \bar{c}_{o\sigma}(\tau)c_{i\sigma}(\tau)\biggr]\\
S^{(o)}=&\int_0^\beta d\tau \biggl[\sum_{i\neq o\,\sigma} \bar{c}_{i\sigma}(\tau)\Bigl(\frac{\partial}{\partial\tau}-\mu\Bigr)c_{i\sigma}(\tau) -\sum_{i\neq o\,j\neq o\,\sigma} t_{ij} \bar{c}_{i\sigma}(\tau)c_{j\sigma}(\tau) \\
&\mspace{60mu}+\frac{U}{2} \sum_{\substack{i\neq o\,\sigma\sigma'\\\sigma\neq\sigma'}} \bar{c}_{i\sigma}(\tau)c_{i\sigma}(\tau)\bar{c}_{i\sigma'}(\tau)c_{i\sigma'}(\tau)\biggr] \notag
\end{align}
The aim is now to integrate out all lattice degrees of freedom except
those of site $o$ in order to find the effective dynamics at site
$o$. In that process, the action $S_o$ remains unchanged, the terms of
$\Delta S$ are expanded in terms of the hopping $t$ which becomes small
with increasing dimension and averaged with respect to the action
$S^{(o)}$. Defining $\Delta S(\tau)$ via $\Delta S=\int_0^\beta d\tau\,\Delta S(\tau)$ the partition
function is
\begin{equation}\begin{split}
Z=\int  \mathcal{D}\bar{c}_{o\sigma}\mathcal{D}c_{o\sigma} e^{-S_o}\int \prod_{i\neq o} \mathcal{D}\bar{c}_{i\sigma}\mathcal{D}c_{i\sigma} e^{-S^{(o)}}e^{-\int_0^\beta d\tau\,\Delta S(\tau)}
\label{eq:partition}\end{split}\end{equation} 
Now we can expand the last exponential function as
\begin{equation}
e^{-\int_0^\beta d\tau\,\Delta S(\tau)}=1-\int_0^\beta d\tau\,\Delta S(\tau)+\frac{1}{2!}\int_0^\beta d\tau_1\int_0^\beta d\tau_2\,\Delta S(\tau_1)\Delta S(\tau_2)-\ldots
\label{eq:expand}\end{equation} 
Taking into account that in general an operator average with respect
to an action $S$ can be expressed as
\begin{equation}
\langle A \rangle_S = \frac{\int \prod_i \mathcal{D}\bar{c}_\alpha\mathcal{D}c_\alpha   e^{-S} A[\bar{c}_\alpha,c_\alpha] }{\int \prod_i \mathcal{D}\bar{c}_\alpha\mathcal{D}c_\alpha   e^{-S}} = Z_s^{-1} \int \prod_i \mathcal{D}\bar{c}_\alpha\mathcal{D}c_\alpha   e^{-S} A[\bar{c}_\alpha,c_\alpha]
\end{equation}
we can consider the second functional integral in \eqref{eq:partition}
to average the terms of the expansion \eqref{eq:expand} with respect
to the lattice action $S^{(o)}$:
\begin{equation}\begin{split}
Z=\int \prod_i \mathcal{D}\bar{c}_{o\sigma}\mathcal{D}c_{o\sigma} e^{-S_o}Z_{S^{(o)}}\biggl\{ &1-\int_0^\beta d\tau\,\langle \Delta S(\tau)\rangle_{S^{(o)}} \\&+\frac{1}{2!}\int_0^\beta d\tau_1\int_0^\beta d\tau_2\,\langle \Delta S(\tau_1)\Delta S(\tau_2)\rangle_{S^{(o)}} -\ldots\biggr\} 
\label{eq:partitionI}\end{split}\end{equation} 
Here, the partition function of the lattice without site $o$ is
abbreviated as
\begin{equation}
Z_{S^{(o)}}=\int \prod_i \mathcal{D}\bar{c}_\alpha\mathcal{D}c_\alpha   e^{-S^{(o)}}\,.
\end{equation}
Now the terms in \eqref{eq:partitionI} with odd powers of $\Delta S$ will
average to zero. For example,
\begin{equation}
\langle \Delta S(\tau)\rangle_{S^{(o)}} = \sum_{i\sigma} t_{io} \langle \bar{c}_{i\sigma}(\tau)\rangle_{S^{(o)}}c_{o\sigma}(\tau) +t_{oi} \bar{c}_{o\sigma}(\tau)\langle c_{i\sigma}(\tau)\rangle_{S^{(o)}} = 0\,,
\end{equation}
because the average $\langle \rangle_{S^{(o)}}$ acts on all sites except $o$.
The next average in \eqref{eq:partitionI} yields
\begin{equation}\begin{split}
&\langle \Delta S(\tau_1)\Delta S(\tau_2)\rangle_{S^{(o)}} = \Bigl\langle T_\tau\biggl[ \sum_{i\sigma} t_{io}  \bar{c}_{i\sigma}(\tau_1)c_{o\sigma}(\tau_1) +t_{oi} \bar{c}_{o\sigma}(\tau_1)c_{i\sigma}(\tau_1)\biggr]\times\\&\mspace{180mu} \times \biggl[ \sum_{j\sigma'} t_{jo}  \bar{c}_{j\sigma'}(\tau_2)c_{o\sigma'}(\tau_2) +t_{oj} \bar{c}_{o\sigma'}(\tau_2)c_{j\sigma'}(\tau_2)\biggr]\Bigr\rangle_{S^{(o)}}\\
&=\sum_{ij\sigma\sigma'} t_{io}t_{oj}  c_{o\sigma}(\tau_1)\langle T_\tau\bar{c}_{i\sigma}(\tau_1)c_{j\sigma'}(\tau_2)\rangle_{S^{(o)}}\;\bar{c}_{o\sigma'}(\tau_2) + \sum_{ij\sigma\sigma'} t_{oi}t_{jo} \bar{c}_{o\sigma}(\tau_1)\langle T_\tau c_{i\sigma}(\tau_1)\bar{c}_{j\sigma'}(\tau_2)\rangle_{S^{(o)}}\;c_{o\sigma'}(\tau_2)\\
&=2\sum_{ij\sigma\sigma'} t_{io}t_{oj}   \bar{c}_{o\sigma}(\tau_1)\langle T_\tau c_{i\sigma}(\tau_1)\bar{c}_{j\sigma'}(\tau_2)\rangle_{S^{(o)}}\;c_{o\sigma'}(\tau_2)\\
&=2\sum_{ij\sigma} t_{io}t_{oj}   \bar{c}_{o\sigma}(\tau_1)\langle T_\tau c_{i\sigma}(\tau_1)\bar{c}_{j\sigma}(\tau_2)\rangle_{S^{(o)}}\;c_{o\sigma}(\tau_2)\\
&=-2\sum_{ij\sigma} t_{io}t_{oj}   \bar{c}_{o\sigma}(\tau_1)G_{ij\,\sigma}^{(o)}(\tau_1-\tau_2)c_{o\sigma}(\tau_2)
\end{split}\end{equation}
The imaginary time ordering operator $T_\tau$ enters because the path
integral leads to imaginary time ordering. Only terms with $\sigma=\sigma'$
contribute as we are considering a paramagnetic state and thus $\langle T_\tau
c_{i\sigma}(\tau_1)\bar{c}_{j\sigma'}(\tau_2)\rangle_{S^{(o)}}=\delta_{\sigma\sigma'}\langle T_\tau
c_{i\sigma}(\tau_1)\bar{c}_{j\sigma}(\tau_2)\rangle_{S^{(o)}}$. We have identified the
average with the cavity Greens function $G_{ij\,\sigma}^{(o)}(\tau_1-\tau_2) =- \langle
T_\tau c_{i\sigma}(\tau_1)c^+_{j\sigma}(\tau_2)\rangle_{S^{(o)}}$, {\it i.$\,$e.} the Greens
function of the Hubbard model without the site $o$.  Now we have for
the partition function
\begin{equation}\begin{split}
Z=&\int \prod_{\sigma} \mathcal{D}\bar{c}_{o\sigma}\mathcal{D}c_{o\sigma} e^{-S_o}Z_{S^{(o)}}\times \\&\times \biggl\{ 1-\int_0^\beta d\tau_1\int_0^\beta d\tau_2\, \sum_{ij\sigma} t_{io}t_{oj}  \bar{c}_{o\sigma}(\tau_1)c_{o\sigma}(\tau_2)G_{ij\,\sigma}^{(o)}(\tau_1-\tau_2)+\ldots\biggr\} 
\label{eq:partitionII}\end{split}\end{equation} 
We would like to write the bracket $\{ \}$ in \eqref{eq:partitionII}
again as an exponential function in order to identify an effective
action $S_{\text{eff}}$:
\begin{equation}
Z=\int \prod_i \mathcal{D}\bar{c}_{o\sigma}\mathcal{D}c_{o\sigma} e^{-S_{\text{eff}}}
\end{equation}
Noting that the next term in the expansion of \eqref{eq:partitionII}
would read
\begin{equation}\begin{split}
\int_0^\beta d\tau_1\int_0^\beta d\tau_2\int_0^\beta d\tau_3\int_0^\beta d\tau_4\, &\sum_{i_1\, i_2\, j_1\, j_2\, \sigma}  \bar{c}_{o\sigma}(\tau_1)\bar{c}_{o\sigma}(\tau_3)c_{o\sigma}(\tau_2)c_{o\sigma}(\tau_4)\times\\&\times t_{i_1\,o}t_{i_2\,o}t_{o\,j_1}t_{o\,j_2}  G_{i_1\,i_2\,j_1\,j_2\,\sigma}^{(o)}(\tau_1\,\tau_3,\tau_2\,\tau_4) 
\end{split}\end{equation}  
we can write for the partition function \eqref{eq:partitionII} 
\begin{equation}\begin{split}
Z=&\int \prod_i \mathcal{D}\bar{c}_{o\sigma}\mathcal{D}c_{o\sigma} e^{-S_o}Z_{S^{(o)}}\times \\&\times 
\exp\biggl\{ -\sum_{n=1}^\infty \sum_\sigma \int_0^\beta d\tau_1  \ldots \int_0^\beta d\tau_{2n} \:\bar{c}_{o\sigma}(\tau_1)\ldots\bar{c}_{o\sigma}(\tau_{2n-1})c_{o\sigma}(\tau_2)\ldots c_{o\sigma}(\tau_{2n}) \times \\&\mspace{50mu}\times \sum_{\substack{i_1,\ldots,i_n\\j_1,\ldots,j_n}} t_{i_1\,o}\ldots t_{i_n\,o}t_{o\,j_1}\ldots t_{o\,j_n} G_{i_1\ldots i_n\,j_1\ldots j_n\,\sigma}^{(o)}(\tau_1\ldots\tau_{2n-1},\tau_2\ldots\tau_{2n}) \biggr\} 
\label{eq:partitionIII}\end{split}\end{equation} 
All terms but the first in this sum over $n$ turn out to be at least
of order $1/d$ so that they vanish in the limit of infinite dimension
$d=\infty$.  Thus, in this limit we find for the effective action
\begin{equation}\begin{split}
S_{\text{eff}} &= S_o + \sum_{\sigma}\int_0^\beta d\tau_1\int_0^\beta d\tau_2\, \bar{c}_{o\sigma}(\tau_1)c_{o\sigma}(\tau_2) \sum_{ij} t_{io}t_{oj}  G_{ij\,\sigma}^{(o)}(\tau_1-\tau_2)\\
&=\int_0^\beta d\tau \biggl[\sum_{\sigma} \bar{c}_{o\sigma}(\tau)\Bigl(\frac{\partial}{\partial\tau}-\mu\Bigr)c_{o\sigma}(\tau)  +\frac{U}{2} \sum_{\substack{\sigma\sigma'\\\sigma\neq\sigma'}} \bar{c}_{o\sigma}(\tau)c_{o\sigma}(\tau)\bar{c}_{o\sigma'}(\tau)c_{o\sigma'}(\tau) \biggr]\\
&\mspace{50mu}+ \sum_{\sigma}\int_0^\beta d\tau_1\int_0^\beta d\tau_2\, \bar{c}_{o\sigma}(\tau_1)c_{o\sigma}(\tau_2) \sum_{ij} t_{io}t_{oj}  G_{ij\,\sigma}^{(o)}(\tau_1-\tau_2)\\
\end{split}\end{equation}
and introducing the Weiss field
\begin{equation}
\mathcal{G}^{-1}_{\sigma}(\tau_1-\tau_2) = -\Bigl(\frac{\partial}{\partial\tau_1}-\mu\Bigr)\delta_{\tau_1\,\tau_2}-\sum_{ij} t_{io}t_{oj}  G_{ij\,\sigma}^{(o)}(\tau_1-\tau_2)
\label{eq:weiss}\end{equation}
we finally get
\begin{equation}
S_{\text{eff}} =-\sum_{\sigma}\int_0^\beta d\tau_1\int_0^\beta d\tau_2\,\bar{c}_{o\sigma}(\tau_1)\mathcal{G}^{-1}_{\sigma}(\tau_1-\tau_2) c_{o\sigma}(\tau_2) +\int_0^\beta d\tau\,\frac{U}{2} \sum_{\substack{\sigma\sigma'\\\sigma\neq\sigma'}} \bar{c}_{o\sigma}(\tau)c_{o\sigma}(\tau)\bar{c}_{o\sigma'}(\tau)c_{o\sigma'}(\tau)
\label{eq:effaction}\end{equation}
The equation
\begin{equation}
G_{ij\,\sigma}^{(o)} = G_{ij\,\sigma} - G_{io\,\sigma}G_{oo\,\sigma}^{-1}G_{oj\,\sigma}
\label{eq:Grelate}\end{equation}
is needed to relate the cavity Greens function to the Greens function
of the lattice $G_{ij\,\sigma}$.  Going from imaginary time to imaginary
frequency and combining with \eqref{eq:Grelate}, the Weiss function
\eqref{eq:weiss} reads
\begin{equation}\begin{split}
\mathcal{G}^{-1}_{\sigma}(i\omega_n) &= i\omega_n+\mu -\sum_{ij} t_{io}t_{oj}  G_{ij\,\sigma}^{(o)}(i\omega_n)\\
&=i\omega_n+\mu -\sum_{ij} t_{io}t_{oj}  \Bigl[ G_{ij\,\sigma}(i\omega_n) - G_{io\,\sigma}(i\omega_n)G_{oo\,\sigma}^{-1}(i\omega_n)G_{oj\,\sigma}(i\omega_n)\Bigr]
\label{eq:weissI}\end{split}\end{equation}
If we now go from real space to $k$ space we can simplify this equation.
Introducing the Fourier transform $G_{k\,\sigma}$ via
\begin{equation}
G_{ij\,\sigma}(i\omega_n)=\sum_k e^{ikR_{ij}}G_{k\,\sigma}(i\omega_n)
\end{equation}
we find
\begin{equation}\begin{split}
\sum_{i} t_{io}G_{io\,\sigma}(i\omega_n)&=\sum_{i} t_{io}\sum_k e^{ikR_{io}}G_{k\,\sigma}(i\omega_n)=\sum_{k}\varepsilon_kG_{k\,\sigma}(i\omega_n)\\
\sum_{ij} t_{io}t_{oj}G_{ij\,\sigma}(i\omega_n)&=\sum_{ij} t_{io}t_{oj}\sum_k e^{ikR_{ij}}G_{k\,\sigma}(i\omega_n)\\
&=\sum_k\sum_it_{io}e^{ikR_{io}} \sum_jt_{oj}e^{ikR_{oj}}G_{k\,\sigma}(i\omega_n)=\sum_{k}\varepsilon_k^2G_{k\,\sigma}(i\omega_n)
\end{split}\end{equation}
In the general form of the Greens function
$G_{k\,\sigma}^{-1}(i\omega_n)=i\omega_n+\mu-\varepsilon_k-\Sigma_\sigma(i\omega_n)$ we introduce the
abbreviation $\xi=i\omega_n+\mu-\Sigma_\sigma(i\omega_n)$ to get $G_{k\,\sigma}^{-1}(i\omega_n)=\xi-\varepsilon_k$
and determine the sums
\begin{equation}\begin{split}
\sum_{k}\varepsilon_kG_{k\,\sigma}(i\omega_n)&=\sum_{k}\frac{\varepsilon_k}{\xi-\varepsilon_k}=\sum_{k}\frac{\varepsilon_k-\xi+\xi}{\xi-\varepsilon_k}=-1+\sum_{k}\frac{\xi}{\xi-\varepsilon_k}\\
&=-1+\xi\sum_{k}G_{k\,\sigma}(i\omega_n)=-1+\xi G_{oo\,\sigma}(i\omega_n)\\
\sum_{k}\varepsilon_k^2G_{k\,\sigma}(i\omega_n)&=\sum_{k}\frac{\varepsilon_k^2}{\xi-\varepsilon_k}=\sum_{k}\frac{\varepsilon_k(\varepsilon_k-\xi)+\varepsilon_k\xi}{\xi-\varepsilon_k}=\sum_{k}\varepsilon_k+\xi\sum_{k}\frac{\varepsilon_k}{\xi-\varepsilon_k}\\
&=\xi\bigl( -1+\xi G_{oo\,\sigma}(i\omega_n)\bigr)=-\xi+\xi^2G_{oo\,\sigma}(i\omega_n)
\end{split}\end{equation}
With this, the Weiss function \eqref{eq:weissI} becomes
\begin{equation}\begin{split}
\mathcal{G}^{-1}_{\sigma}(i\omega_n) &=i\omega_n+\mu -\sum_{k}\varepsilon_k^2G_{k\,\sigma}(i\omega_n) + \Bigl(\sum_{k}\varepsilon_kG_{k\,\sigma}(i\omega_n)\Bigr)^2G_{oo\,\sigma}^{-1}(i\omega_n)\\
&=i\omega_n+\mu+\xi-\xi^2G_{oo\,\sigma}(i\omega_n)+\bigl(-1+\xi G_{oo\,\sigma}(i\omega_n)\bigr)\bigl(-G_{oo\,\sigma}^{-1}(i\omega_n)+\xi\bigr)\\
&=i\omega_n+\mu-\xi+G_{oo\,\sigma}^{-1}(i\omega_n)=\Sigma_\sigma(i\omega_n)+G_{oo\,\sigma}^{-1}(i\omega_n)
\label{eq:weissII}\end{split}\end{equation}
This equation
$G_{oo\,\sigma}^{-1}(i\omega_n)=\mathcal{G}^{-1}_{\sigma}(i\omega_n)-\Sigma_\sigma(i\omega_n)$ is the
Dyson equation for the local Greens function.

The effective action \eqref{eq:effaction} can now be interpreted in
terms of the Anderson impurity model, {\it i.$\,$e.} the Anderson
impurity model gives rise to an action which becomes identical to
\eqref{eq:effaction} if an additional self consistency condition is
fulfilled.  The Hamiltonian for the Anderson impurity model is
\begin{equation}
H=\sum_{k\sigma}\varepsilon_kc^+_{k\sigma}c_{k\sigma} +\sum_{k\sigma}\bigl(V_kc^+_{k\sigma}f_\sigma +V^*_kf^+_\sigma c_{k\sigma}\bigr) - \sum_\sigma \mu f^+_\sigma f_\sigma + \frac{U}{2}  \sum_{\substack{\sigma\sigma'\\\sigma\neq\sigma'}}f^+_\sigma f_\sigma f^+_{\sigma'}f_{\sigma'}
\label{eq:AIMhamiltonian}\end{equation}
where $\sigma$ runs from 1 to the degeneracy $N$. The action corresponding
to this Hamiltonian will consist of a purely local part $S_o$
concerning only the $f$ electrons
\begin{equation}
S_o=\int_0^\beta d\tau \biggl[\sum_{\sigma} \bar{f}_{\sigma}(\tau)\Bigl(\frac{\partial}{\partial\tau}-\mu\Bigr)f_{\sigma}(\tau)+ \frac{U}{2}  \sum_{\substack{\sigma\sigma'\\\sigma\neq\sigma'}}\bar{f}_\sigma(\tau)f_\sigma(\tau)\bar{f}_{\sigma'}(\tau)f_{\sigma'}(\tau)\biggr]
\end{equation}
 and a part involving conduction band electrons that can be integrated
 out:
\begin{equation}\begin{split}
S=S_o+\int_0^\beta d\tau \sum_{k\sigma}\biggl[& \bar{c}_{k\sigma}(\tau)\Bigl(\frac{\partial}{\partial\tau}+\varepsilon_k\Bigr)c_{k\sigma}(\tau) %\\&
+ V_k\bar{c}_{k\sigma}(\tau)f_\sigma(\tau) +V^*_k\bar{f}_\sigma(\tau)c_{k\sigma}(\tau)\biggr]
\end{split}\end{equation}
Now the partition function for the Hamiltonian \eqref{eq:AIMhamiltonian} is
\begin{equation}\begin{split}
 Z&=\int \mathcal{D}\bar{f}_{\sigma}\mathcal{D}f_{\sigma} \int \prod_i  \mathcal{D}\bar{c}_{i\sigma}\mathcal{D}c_{i\sigma}  e^{-S}%\\&
=\int \mathcal{D}\bar{f}_{\sigma}\mathcal{D}f_{\sigma} \,e^{-S_o} \int \prod_i  \mathcal{D}\bar{c}_{i\sigma}\mathcal{D}c_{i\sigma} \times\\&\times\exp\Biggl\{  \int_0^\beta d\tau \sum_{k\sigma}\biggl[ \bar{c}_{k\sigma}(\tau)\Bigl(\frac{\partial}{\partial\tau}+\varepsilon_k\Bigr)c_{k\sigma}(\tau) 
+ V_k\bar{c}_{k\sigma}(\tau)f_\sigma(\tau) +V^*_k\bar{f}_\sigma(\tau)c_{k\sigma}(\tau)\biggr] \Biggr\} \\
&=\int \mathcal{D}\bar{f}_{\sigma}\mathcal{D}f_{\sigma} \,e^{-S_o} \prod_k\det\Bigl(\frac{\partial}{\partial\tau}+\varepsilon_k\Bigr)\times\\&\times\exp\biggl\{ \sum_{k\sigma}\int_0^\beta d\tau_1\int_0^\beta d\tau_2\, \bar{f}_\sigma(\tau_1)V^*_k V_k\Bigl(\frac{\partial}{\partial\tau_1}+\varepsilon_k\Bigr)^{-1}\delta_{\tau_1\,\tau_2}f_\sigma(\tau_2)\biggr\} 
\end{split}\end{equation}
In the last step, the terms involving $f$ electrons
$V^*_k\bar{f}_\sigma(\tau)$ and $V_kf_\sigma(\tau)$ were taken as source terms, which
makes the term in the exponent a Gaussian integral that can be
evaluated directly. The determinant constitutes a constant factor in
the partition function that doesn't concern us here. We are left with
an action for the $f$ electrons that reads
\begin{equation}\begin{split}
S_f=& \int_0^\beta d\tau_1\int_0^\beta d\tau_2 \sum_{\sigma} \bar{f}_{\sigma}(\tau_1)\biggl[\Bigl(\frac{\partial}{\partial\tau_1}-\mu\Bigr)\delta_{\tau_1\,\tau_2} -\sum_k|V_k|^2\Bigl(\frac{\partial}{\partial\tau_1}+\varepsilon_k\Bigr)^{-1}\delta_{\tau_1\,\tau_2} \biggr] f_{\sigma}(\tau_2)
\\&+ \int_0^\beta d\tau\frac{U}{2}  \sum_{\substack{\sigma\sigma'\\\sigma\neq\sigma'}}\bar{f}_\sigma(\tau)f_\sigma(\tau)\bar{f}_{\sigma'}(\tau)f_{\sigma'}(\tau)
\end{split}\end{equation}
If we now compare this to the effective action of the Hubbard model
\eqref{eq:effaction}, we see that they are identical if we require
that the Weiss function $\mathcal{G}(\tau_1-\tau_2)$ fulfils the condition
\begin{equation}
\mathcal{G}^{-1}(\tau_1-\tau_2)=-\Bigl(\frac{\partial}{\partial\tau_1}-\mu\Bigr)\delta_{\tau_1\,\tau_2} +\sum_k|V_k|^2\Bigl(\frac{\partial}{\partial\tau_1}+\varepsilon_k\Bigr)^{-1}\delta_{\tau_1\,\tau_2}
\end{equation}
Going from imaginary time to imaginary frequency, this equation reads
\begin{equation}
\mathcal{G}^{-1}(i\omega_n)=i\omega_n+\mu -\sum_k\frac{|V_k|^2}{i\omega_n-\varepsilon_k}
\label{eq:weissAIM}\end{equation}
Here we can identify the usual definition of the hybridization
function $\Delta(i\omega_n)$ in the Anderson impurity model
\begin{equation}
\Delta(i\omega_n)=\sum_k\frac{|V_k|^2}{i\omega_n-\varepsilon_k}
\end{equation}
If we now equate Weiss functions \eqref{eq:weissII} and
\eqref{eq:weissAIM} we find the DMFT self-consistency condition in
terms of a prescription for $\Delta(i\omega_n)$
\begin{equation}
\Delta(i\omega_n)=i\omega_n+\mu-\Sigma_\sigma(i\omega_n)-G_{oo\,\sigma}^{-1}(i\omega_n)
\label{eq:selfconsistency}\end{equation}
On the Bethe lattice and with a half band width of $D=2t$, we have a
noninteracting density of states
\begin{equation}
\rho_0(\varepsilon) = \frac{1}{2\pi t^2} \sqrt{4t^2-\varepsilon^2}
\end{equation}
and thus we can write for the local Greens function
\begin{equation}\begin{split}
G_{oo\,\sigma}(i\omega_n) &= \sum_k G_k(i\omega_n)= \sum_k \frac{1}{\xi-\varepsilon_k}\qquad \text{with}\quad\xi=i\omega_n+\mu-\Sigma_\sigma(i\omega_n)\\&= \int d\varepsilon\frac{\rho_0(\varepsilon)}{\xi-\varepsilon}=\frac{1}{2\pi t^2}\int_{-2t}^{2t} d\varepsilon\frac{\sqrt{4t^2-\varepsilon^2}}{\xi-\varepsilon}=\frac{1}{2 t^2}\bigl(\xi-\text{sgn}(\text{Re}\,\xi)\sqrt{\xi^2-4t^2}\bigr)
\end{split}\end{equation}
From this we gain the expression
\begin{equation}
t^2G_{oo\,\sigma}(i\omega_n)-\xi+G_{oo\,\sigma}^{-1}(i\omega_n)=0\,,
\end{equation}
which combined with Eq.~\eqref{eq:selfconsistency} leads to a
simplified form of the self-consistency condition
\begin{equation}
\Delta(i\omega_n)=t^2G_{oo\,\sigma}(i\omega_n)\,.
\label{eq:selfconsistencyI}\end{equation}

\section{DMFT self consistency condition for the Anderson lattice}\label{app:and}

We again focus on one site $i=o$ and split the Hamiltonian into three
parts:
\begin{align}
H_{\text{PAM}}=&H_o + H_c + H^{(o)}\\
H_o=& \varepsilon_c \sum_{\sigma} c^+_{o\sigma}c_{o\sigma}+\varepsilon_f \sum_{\sigma} f^+_{o\sigma}f_{o\sigma}\notag\\&+\sum_{\sigma}\bigl(V_{o\sigma}c^+_{o\sigma}f_{o\sigma} +V^*_{o\sigma}f^+_{o\sigma}c_{o\sigma}\bigr)+\frac{U}{2} \sum_{\substack{\sigma\sigma'\\\sigma\neq\sigma'}}f^+_{o\sigma}f_{o\sigma}f^+_{o\sigma'}f_{o\sigma'}\\
H_c=&-\sum_{i\sigma} \Bigl[ t^c_{io} c^+_{i\sigma}c_{o\sigma} +t^c_{oi} c^+_{o\sigma}c_{i\sigma}\Bigr]\\%-\sum_{i\sigma} \Bigl[ t^f_{io} f^+_{i\sigma}f_{o\sigma} +t^f_{oi} f^+_{o\sigma}f_{i\sigma}\Bigr]\\
H^{(o)}=& -\sum_{i\neq o\,j\neq o\,\sigma} t^c_{ij} c^+_{i\sigma}c_{j\sigma}%-\sum_{i\neq o\,j\neq o\,\sigma} t^f_{ij} f^+_{i\sigma}f_{j\sigma}
+\varepsilon_c \sum_{i\neq o\,\sigma} c^+_{i\sigma}c_{i\sigma} +\varepsilon_f \sum_{i\neq o\,\sigma} f^+_{i\sigma}f_{i\sigma} \notag\\&+\sum_{i\neq o\,\sigma}\bigl(V_{i\sigma}c^+_{i\sigma}f_{i\sigma} +V^*_{i\sigma}f^+_{i\sigma}c_{i\sigma}\bigr)+\frac{U}{2} \sum_{\substack{i\neq o\,\sigma\sigma'\\\sigma\neq\sigma'}}f^+_{i\sigma}f_{i\sigma}f^+_{i\sigma'}f_{i\sigma'}
\end{align}
$H_c$ has the same form as in the Hubbard model, but the local part
$H_o$ is more complicated as it contains two species of electrons,
conduction and $f$ electrons.  Nevertheless, we can proceed completely
along the lines detailed for the Hubbard model above, expanding the
action $\Delta S$ arising from $H_c$ in order to arrive at an effective
action for site $o$. In this case we have
\begin{equation}\begin{split}
S_o=& \int_0^\beta d\tau \biggl[\sum_{\sigma} \bar{f}_{\sigma}(\tau)\Bigl(\frac{\partial}{\partial\tau}+\varepsilon_f\Bigr)f_{\sigma}(\tau)+ \frac{U}{2}  \sum_{\substack{\sigma\sigma'\\\sigma\neq\sigma'}}\bar{f}_{o\sigma}(\tau)f_{o\sigma}(\tau)\bar{f}_{o\sigma'}(\tau)f_{o\sigma'}(\tau)\\&+\sum_{\sigma} \bar{c}_{o\sigma}(\tau)\Bigl(\frac{\partial}{\partial\tau}+\varepsilon_c\Bigr)c_{o\sigma}(\tau)+\sum_{\sigma} \bigl(V_{o\sigma}\bar{c}_{o\sigma}(\tau)f_{o\sigma}(\tau) +V^*_{o\sigma}\bar{f}_{o\sigma}(\tau)c_{o\sigma}(\tau)\bigr)\biggr]
\end{split}\end{equation}
and
\begin{equation}\begin{split}
S_{\text{eff}} &= S_o + \int_0^\beta d\tau_1\int_0^\beta d\tau_2\, \bar{c}_{o\sigma}(\tau_1)c_{o\sigma}(\tau_2) \sum_{ij\sigma} t_{io}t_{oj}  G_{ij\,\sigma}^{(o)}(\tau_1-\tau_2)
\end{split}\end{equation}

In the $d\to \infty$ limit, the Green's function becomes
\begin{equation}
G^{-1}(i\omega_n,\mathbf{k}) = \begin{pmatrix}i\omega_n+\mu-\varepsilon_f^0-\Sigma_f(i\omega_n)&V_k\\V_k&i\omega_n+\mu-\varepsilon_k\end{pmatrix}
\end{equation}
Inverting the matrix according to
\begin{equation}
M=\begin{pmatrix}A&B\\B&C\end{pmatrix}\curvearrowright M^{-1}=\frac{1}{\text{det} M}\begin{pmatrix}C&-B\\-B&A\end{pmatrix}
\end{equation}
we find
\begin{equation}\begin{split}
G(i\omega_n,\mathbf{k}) =& \frac{1}{(i\omega_n+\mu-\varepsilon_f^0-\Sigma_f(i\omega_n))(i\omega_n+\mu-\varepsilon_k)-V_k^2}\times \\&\times \begin{pmatrix}i\omega_n+\mu-\varepsilon_k&-V_k\\-V_k&i\omega_n+\mu-\varepsilon_f^0-\Sigma_f(i\omega_n)\end{pmatrix}
\end{split}\end{equation}
Thus, we find for the $f$ electron Green's function
\begin{equation}\begin{split}
G_f(i\omega_n,\mathbf{k}) = \biggl(i\omega_n+\mu-\varepsilon_f^0-\Sigma_f(i\omega_n) -\frac{V_k^2}{i\omega_n+\mu-\varepsilon_k}\biggr)^{-1}
\end{split}\end{equation}
and for the conduction band Green's function
\begin{equation}\begin{split}
G_c(i\omega_n,\mathbf{k}) = \biggl( i\omega_n+\mu-\varepsilon_k-\frac{V_k^2}{i\omega_n+\mu-\varepsilon_f^0-\Sigma_f(i\omega_n)}\biggr)^{-1}
\end{split}\end{equation}
We get the local propagators as $G_f(R=0,i\omega_n)=\sum_{\mathbf{k}}
G_f(i\omega_n,\mathbf{k})e^{ik(R=0)}$ by summation over $\mathbf{k}$:
\begin{equation}\begin{split}
G_f^{\text{local}}(i\omega_n) &= \sum_{\mathbf{k}} G_f(i\omega_n,\mathbf{k}) \\&
= \int d\varepsilon \,\rho_0(\varepsilon) \biggl(i\omega_n+\mu-\varepsilon_f^0-\Sigma_f(i\omega_n) -\frac{V(\varepsilon)^2}{i\omega_n+\mu-\varepsilon}\biggr)^{-1}\\
G_c^{\text{local}}(i\omega_n) &= 
\int d\varepsilon \,\rho_0(\varepsilon) \biggl( i\omega_n+\mu-\varepsilon-\frac{V(\varepsilon)^2}{i\omega_n+\mu-\varepsilon_f^0-\Sigma_f(i\omega_n)}\biggr)^{-1}
\label{eq:Glocal}\end{split}\end{equation}
For computational purposes it is useful to note that for the case of
an energy independent $V(\varepsilon)\equiv V$, $G_c^{\text{local}}(i\omega_n)$ can be
written as a Hilbert transform $\tilde{D}(\zeta)=\int_{-\infty}^{\infty}d\varepsilon
\frac{D(\varepsilon)}{\zeta-\varepsilon}$:
\begin{equation}\begin{split}
G_c^{\text{local}}(i\omega_n) &= \tilde{D}\biggl( i\omega_n+\mu -\frac{V^2}{i\omega_n+\mu-\varepsilon_f^0-\Sigma_f(i\omega_n)} \biggr)
\label{eq:Gc}\end{split}\end{equation}
Rewriting $G_f^{\text{local}}(i\omega_n)$, we can likewise reduce the
energy integral to the calculation of a Hilbert transform:
\begin{equation}\begin{split}
G_f^{\text{local}}(i\omega_n) &= 
 \int d\varepsilon \,\rho_0(\varepsilon) \biggl\{ \frac{1}{i\omega_n+\mu-\varepsilon_f^0-\Sigma_f(i\omega_n)}\\
&+\frac{V^2}{\bigl(i\omega_n+\mu-\varepsilon_f^0-\Sigma_f(i\omega_n)\bigr)^2}\frac{1}{i\omega_n+\mu-\varepsilon-\frac{V^2}{i\omega_n+\mu-\varepsilon_f^0-\Sigma_f(i\omega_n)}}  \biggr\} 
\label{eq:Gf1}\end{split}\end{equation}
and with Eq.~\eqref{eq:Gc}
\begin{equation}\begin{split}
G_f^{\text{local}}(i\omega_n)&=\frac{1}{i\omega_n+\mu-\varepsilon_f^0-\Sigma_f(i\omega_n)}+\frac{V^2}{\bigl(i\omega_n+\mu-\varepsilon_f^0-\Sigma_f(i\omega_n)\bigr)^2}G_c^{\text{local}}(i\omega_n)
\label{eq:Gf2}\end{split}\end{equation}
If we now assume a semicircular DOS $D(\varepsilon)=\frac{1}{2\pi
t^2}\sqrt{4t^2-\varepsilon^2}$ for the hybridization $V_k$ we can explicitly
write for the Hilbert transform
\begin{equation}\begin{split}
&\tilde{D}(\zeta)=\frac{1}{2\pi t^2}\int_{-2t}^{2t}d\varepsilon \frac{\sqrt{4t^2-\varepsilon^2}}{\zeta-\varepsilon}=\frac{1}{2 t^2}\bigl(\zeta-\text{sgn}(\text{Re}\,\zeta)\sqrt{\zeta^2-4t^2}\bigr)
\label{eq:hilbert}\end{split}\end{equation}
Thus, on the Bethe lattice the self consistency condition can be
calculated without an integral over energies.  We also need the Dyson
equation
\begin{equation}
\mathcal{G}_0^{-1}(i\omega_n)=G_f^{-1}(i\omega_n)+\Sigma(i\omega_n)
\label{eq:dyson}\end{equation}
From the high frequency limit of this equation we can find the form of
the Weiss function $\mathcal{G}_0^{-1}(i\omega_n)$ by comparing the terms
of the expansion order by order. Expanding Eq.~\eqref{eq:Glocal} we
find
\begin{equation}
G_f(i\omega_n)\approx \frac{1}{i\omega_n}+(\varepsilon_f^0-\mu+\Sigma_f(i\omega_n))\Bigl(\frac{1}{i\omega_n}\Bigr)^2 \quad \text{for }i\omega_n\to\infty 
\end{equation}
Expanding the inverse, we find
\begin{equation}
G_f^{-1}(i\omega_n)\approx i\omega_n+\mu-\varepsilon_f^0-\Sigma_f(i\omega_n)  \quad \text{for }i\omega_n\to\infty 
\end{equation}
Thus, we find from Eq.~\eqref{eq:dyson} the high frequency form of the
Weiss function:
\begin{equation}
\mathcal{G}_0^{-1}(i\omega_n)\approx i\omega_n+\mu-\varepsilon_f^0
\label{eq:highfreq_weiss}\end{equation}
The hybridization function $\Delta(i\omega_n)$ contains what we have neglected
in the high frequency expansion:
\begin{equation}
\mathcal{G}_0^{-1}(i\omega_n)=i\omega_n+\mu-\varepsilon_f^0-\Delta(i\omega_n)
\label{eq:weissPAM}\end{equation}

\end{appendix}

\newpage

\end{document}